\documentclass[11pt,a4paper]{article} 
\usepackage{jheppub}
\usepackage{enumerate}
\usepackage{wasysym} 
\usepackage{mathrsfs} 
\usepackage{graphicx} 
\usepackage{amsfonts} 
\usepackage{amsbsy} 
\usepackage{amscd}
\usepackage{slashed} 
\usepackage{multirow} 
\usepackage{cancel}
\usepackage{xcolor}
\usepackage{orcidlink}

\makeatletter

\makeatletter

\newcommand{\fmslash}[2][0mu]{%
  \mathchoice
    {\fmsl@sh\displaystyle{#1}{#2}}%
    {\fmsl@sh\textstyle{#1}{#2}}%
    {\fmsl@sh\scriptstyle{#1}{#2}}%
    {\fmsl@sh\scriptscriptstyle{#1}{#2}}}
\newcommand{\fmsl@sh}[3]{%
  \m@th\ooalign{$\hfil#1\mkern#2/\hfil$\crcr$#1#3$}}
\makeatother

\newcommand{\lsim}{{\;\raise0.3ex\hbox{$<$\kern-0.75em\raise-1.1ex\hbox{$\sim$}}\;}}
\newcommand{\gsim}{{\;\raise0.3ex\hbox{$>$\kern-0.75em\raise-1.1ex\hbox{$\sim$}}\;}}

\newcommand{\D}{\fmslash D}

\usepackage{array}
\newcolumntype{C}[1]{>{\centering\arraybackslash$}p{#1}<{$}}

\usepackage{bm} 
\newcommand{\be}{\begin{equation}}
\newcommand{\ee}{\end{equation}}
\newcommand{\bes}{\begin{equation*}}
\newcommand{\ees}{\end{equation*}}
\newcommand{\bea}{\begin{eqnarray}}
\newcommand{\eea}{\end{eqnarray}}
\newcommand{\beas}{\begin{eqnarray*}}
\newcommand{\eeas}{\end{eqnarray*}}
\title{Dark Matter Freeze-in from a $Z^\prime$ Reheaton}

\author[1]{Avirup Ghosh\,\,\orcidlink{0000-0002-4781-842X},}
\author[1]{Alexei H. Sopov\,
\orcidlink{0000-0003-1094-6513},}
\author[1]{and Raymond R. Volkas\,\,\orcidlink{0000-0002-4254-8520}}
\affiliation[1]{ARC Centre of Excellence for Dark Matter Particle Physics, School of Physics, The University of Melbourne, Victoria 3010, Australia}

\emailAdd{avirup.ghosh@unimelb.edu.au}
\emailAdd{sopova@student.unimelb.edu.au}
\emailAdd{raymondv@unimelb.edu.au}

\abstract{
We consider the Standard Model (SM) extended by a secluded $U(1)_D$ gauge sector encompassing Dirac fermion ($\chi$) dark matter (DM), an abelian gauge boson $Z^\prime$ and a SM-singlet complex-scalar field $\Phi$, whose radial component drives cosmic inflation. When the Higgs portal coupling is small, the $Z^\prime$ then acts as a \textit{``reheaton''}, dominating the energy budget of the Universe before finally yielding the SM bath, with reheating temperature $< O(10)$ TeV, through the gauge portal interaction. We explore the possibility that DM freezes-in via non-thermal $Z^\prime$ decays before reheating ends, giving rise to substantial viable parameter space. We account for non-perturbative effects, relevant during the initial stages of reheating, using lattice simulations. We additionally show how the cosmological gravitational wave (GW) background produced by preheating and inflation allow for a direct probe of the reheating mechanism.
}

\begin{document} 

\maketitle
\section{Introduction}
\label{sec:sec1}

Although dark matter (DM) constitutes nearly $27\%$ of the total 
energy density of the Universe~\cite{Planck:2018vyg}, 
till now all evidence for it is purely gravitational. Its non-gravitational 
properties, viz., the DM particle spin, Standard Model (SM) 
weak charges and interactions with other SM particles 
are still unknown despite the combined efforts of direct, 
indirect and collider search experiments to unveil them. 
Given the strong constraints on the DM-SM couplings, 
several scenarios that have been proposed to explain the 
nature of DM feature a beyond-SM (BSM) particle that acts as a 
phenomenologically acceptable portal between DM and SM particles. 
Among these, in the case of fermionic DM candidates, 
is the renormalizable vector boson portal interaction which arises 
in secluded $U(1)_D$ models (see, for example, ~\cite{Pospelov:2008zw}). 
These models augment the fermionic DM particle ($\chi$) 
with a vector boson $Z^\prime$ which kinetically mixes with the 
SM hypercharge gauge boson $B$. In this context, several studies 
have considered the production of thermal dark matter via 
annihilation through $Z^\prime$-mediated 
interactions~\cite{Chun:2010ve,Gondolo:2011eq} 
as well as non-thermal freeze-in production from a thermal bath of 
$Z^\prime$ bosons (for a representative list see~\cite{Cosme:2021baj,Arcadi:2024obp,Barman:2024lxy}).\footnote{A $Z^\prime$ mediator has also been recently studied in the context of (ultra-relativistic) freeze-out of dark matter during reheating~\cite{Henrich:2025pca,Henrich:2025sli}.}

Observationally, this secluded $Z^\prime$ must have a mass, which in a 
renormalizable model may be generated either via the Stueckelberg 
mechanism or via the spontaneous breaking of the secluded $U(1)_D$;
we adopt the latter here.\footnote{Although the presence of vector 
boson mass term does not spoil the renormalizability of the Stueckelberg 
action, the addition of any new interactions can lead to some 
inconsistencies (see, e.g.,~\cite{Ahmed:2020fhc,Kolb:2020fwh,Kribs:2022gri}).}
The spontaneous breaking of the secluded $U(1)_D$ requires 
the presence of a $U(1)_D$-charged dark-scalar 
$\Phi$. The inclusion of $\Phi$ raises the possibility 
of realising cosmological inflation within this 
framework~\cite{Khan:2023uii,Khan:2025keb}. 
While, within the purview of this type of scenarios, 
Refs.~\cite{Khan:2023uii,Khan:2025keb} considered the $Z^\prime$ 
to have a thermal abundance in the post-inflationary radiation 
dominated Universe, we show that an interesting alternative cosmology 
is in principle possible within this set-up given the 
intricacies involved in the transition of the inflaton-dominated 
Universe to the SM radiation-dominated era~\cite{Kofman:1994rk}. 

Specifically, we point out that, analogously to the production of 
SM weak bosons via preheating after Higgs Inflation~\cite{Garcia-Bellido:2008ycs, Bezrukov:2008ut, Repond:2016sol}, 
here also the $U(1)_D$ gauge bosons $(Z^\prime)$ are copiously produced during the 
preheating era provided the vacuum expectation value (VEV) of 
$\Phi$ is much smaller than its field value at the end of 
inflation~\cite{Antusch:2021aiw}.\footnote{Other works which 
have studied the production of vector 
bosons during preheating have considered the vector to be the DM 
candidate and hence very light (see, for 
example~\cite{Graham:2015rva,Dror:2018pdh}).}
Taking the SM Higgs ($h$)-dark scalar ($\Phi$) quartic coupling 
to be minuscule and the inflaton to be heavier than the $Z^\prime$, 
we find that a bath of longitudinal $Z^\prime$ and radial $\Phi$ bosons is 
formed by the end of the preheating era, a process we study using lattice simulations. Later, when $\Phi$ settles 
to its minimum, it dominantly decays into $Z^\prime$ states thereby 
producing a non-thermal $Z^\prime$-dominated era that persists 
until $Z^\prime$ decays lead to the production of SM particles, 
and gives rise to a standard radiation dominated era prior 
to Big-Bang Nucleosynthesis (BBN). Hence, we dub $Z^\prime$ to be a 
``\textit{reheaton}", a role which till now has only been studied considering 
spin-0 bosonic fields~\cite{Opferkuch:2019zbd,Bettoni:2021zhq,Ghoshal:2022ruy,Laverda:2023uqv,Laverda:2024qjt,Figueroa:2024asq,Laverda:2025pmg,Bernal:2025lxp}. 

The duration of the non-thermal $Z^\prime$-dominated era depends 
on the value of both the kinetic mixing parameter $\epsilon$ and the 
$Z^\prime$ mass ($m_{Z^\prime,0}$). The parameter $\epsilon$ (for any given $m_{Z^\prime,0}$) 
is bounded from above by the requirement of keeping the $Z^\prime$ bosons 
non-thermal during the entire cosmological evolution.\footnote{Above this 
critical value of $\epsilon$, $Z^\prime$ states produced from 
inflaton oscillations thermalise with the SM. The cosmology of such a 
scenario is rather straightforward and quite well-understood 
(see~\cite{Khan:2023uii,Khan:2025keb,Cosme:2021baj,Arcadi:2024obp}).}  
But there is also a lower bound on $\epsilon$ (and $m_{Z^\prime,0}$), 
which comes from the requirement that a SM dominated radiation era is 
realised prior to BBN. We also find that these non-thermal $Z^\prime$ 
bosons are non-relativistic for a substantial part of the cosmological 
evolution and provide a natural way to realise intermediate matter 
domination (IMD) prior to SM reheating. 

Furthermore, we find that the DM candidate $\chi$ can be sufficiently 
produced during reheating, primarily from $Z^\prime$ decays, 
provided that the $\chi$ mass is smaller than half of the $Z^\prime$ mass. 
Due to the smallness of $\epsilon$ we find that the DM $\chi$ always remains 
non-thermal and acts as a freeze-in DM candidate. Interestingly, 
for sub-GeV $Z^\prime$ bosons, a substantial region of the parameter space 
is found to be ruled out by several existing observations, while some 
currently allowed regions may be probed by the upcoming SHiP~\cite{SHiP:2020vbd} 
experiment. 

We also find that the non-trivial duration of the matter-dominated 
era in our scenario modifies the stochastic background of gravitational 
waves (SGWB) in a way such that  the present day spectrum lies 
within the detectability range of future GW searches such as the 
\textit{Deci-Hertz Interferometer Gravitational-Wave Observatory} 
(DECIGO~\cite{Seto:2001qf}, specifically for its ``ultimate" sensitivity reach~\cite{Kuroyanagi_2015}), 
a proposed $\mu$-mHz observatory such as $\mu$Ares~\cite{Sesana_2021} if its 
sensitivity could be enhanced by a few orders, or potentially 
by the \textit{Big-Bang Observer} (BBO~\cite{Crowder:2005nr}) for 
certain allowed choices of the model parameters.

The remainder of this paper is organised as follows: In Sec.~\ref{sec:sec2}, 
we discuss the secluded gauged $U(1)_D$ model in detail. 
Section~\ref{sec:sec3} is devoted to discussions of the inflationary 
dynamics, $Z^\prime$ production during inflation and the non-perturbative 
preheating epoch, followed by the perturbative epoch of DM 
and SM production. In Section~\ref{sec:sec4}, we point 
out the possible relic GW spectrum of this scenario. 
Finally, in Section~\ref{sec:sec5}, we summarise and conclude. 
Several formulae relevant for our present study are also 
presented in appendices~\ref{sec:pertdecay} 
and~\ref{sec:thermalization}.

\section{The Model}
\label{sec:sec2}

As mentioned in the introduction, we consider the SM extended 
by a secluded $U(1)_D$ gauge group under which a complex scalar 
field $\Phi$ has charge +1. The associated abelian gauge boson 
is denoted by $Z^\prime$. We also include a Dirac 
fermion $\chi$ as the DM candidate. It has a $U(1)_D$ gauge charge 
$Q_\chi$ and a Dirac mass $m_\chi$. 

The action for this SM extension contains
the additional terms 
\begin{eqnarray}
S &=& \int d^4x \sqrt{-g}\bigg[ -\frac{M^2_P}{2} \left(1+\frac{\xi_\Phi\,\Phi^\dagger \Phi}{M^2_P}+\frac{\xi_H\,H^\dagger H}{M^2_P}\right) R + \left(\tilde{D}_\mu \Phi\right)^\dagger \left(\tilde{D}^\mu \Phi\right) +  \left(D_\mu H\right)^\dagger \left(D^\mu H\right)\nonumber\\
&& \hspace{2.1cm} - V(\Phi,H) - \frac{1}{4}F^\prime_{\mu\nu}F^{\prime \mu\nu} - \frac{\epsilon}{2}F^\prime_{\mu\nu} B^{\mu\nu}_Y + \bar{\chi}i \D \chi - m_\chi\bar{\chi}\chi\bigg],
\label{eq:modelaction}
\end{eqnarray}
where, $g$ is the determinant of the space-time metric, which we take to be (spatially flat) Friedmann-Lema\^itre-Robertson-Walker (FLRW) 
\begin{equation}
    g_{\mu\nu}\mathrm{d}x^\mu\mathrm{d}x^\nu = \mathrm{d}t^2 - a^2(t)\mathrm{d}x^i\mathrm{d}x^i = a^2(\tau)[\mathrm{d}\tau^2 -\mathrm{d}x^i\mathrm{d}x^i],
\end{equation}
where $\tau$ is the conformal time (we denote $' := \frac{\mathrm{d}}{\mathrm{d}\tau}$) and $a$ is the scale factor; $M_P$ is the reduced Planck mass;
$R$ is the 
Ricci Scalar; 
$\tilde{D}_\mu \Phi = \partial_\mu \Phi - i\,g_D\,Z^{\prime}_{\mu}\Phi$ 
and $D_\mu \chi = \partial_\mu \chi  - i\,g_D\,Q_\chi\,Z^{\prime}_{\mu} \chi$ 
are gauge-covariant derivatives with $g_D$ being the $U(1)_D$ 
gauge coupling constant; 
$F^\prime_{\mu\nu} = \partial_\mu Z^\prime_\nu - \partial_\nu Z^\prime_\mu$ is 
the $U(1)_D$ field-strength tensor; and the scalar potential is given by
\begin{eqnarray}
V(\Phi,H) &=& \frac{\lambda_H}{2} \left( H^\dagger H - \frac{v^2_H}{2}\right)^2 + \frac{\lambda_\Phi}{2} \left( \Phi^\dagger \Phi - \frac{v^2_D}{2}\right)^2 + \lambda_{\Phi H} \left( H^\dagger H - \frac{v^2_H}{2}\right) \left( \Phi^\dagger \Phi - \frac{v^2_D}{2}\right).\nonumber\\   
\label{eq:modelscalarpotential}
\end{eqnarray}
In order to realise inflation, we assume non-zero values for the dimensionless non-minimal gravitational coupling parameters $\xi_{\Phi}$ and $\xi_{H}$ (technically required in curved spacetime~\cite{Birrell:1982ix}), leading to additional terms invariant under the extended gauge group.

For $\lambda_H$, $\lambda_\Phi > 0$ and $\lambda_{H\Phi} < \lambda_H\lambda_\Phi$, the scalar potential is minimised when $|\Phi|=v_D/\sqrt{2}$ and $|H|=v_H/\sqrt{2}$. In unitary gauge, the phase component of $\Phi$ is eaten to give rise to a longitudinal polarisation for the $Z'$, which acquires a mass $m_{Z^\prime,0} \equiv g_D v_D$, while the radial component is a dark Higgs, with mass $m_{\phi,0} \simeq \sqrt{\lambda_\Phi} v_D$, which mixes with the SM Higgs $h$ due to the presence of the $\lambda_{\Phi H}$ term in the scalar potential. This produces a mixing angle,
\begin{eqnarray}
\sin\theta &=& \frac{\lambda_{\Phi H}v_H v_D}{\sqrt{(\lambda_\Phi v^2_D - \lambda_H v^2_H)^2 + (\lambda_{\Phi H}v_H v_D)^2}}, 
\label{eq:mixangle}
\end{eqnarray}
for the two real scalar mass eigenstates, which is insignificant for $v_H \ll v_D$ and $|\lambda_{\Phi H}| < \lambda_\Phi$.

The dark sector parameters $v_D$, $\epsilon$, $m_\chi$ and $Q_\chi$ are taken as free in principle, but are constrained by the consistent implementation of freeze-in and perturbative reheating (see Figure \ref{fig:relicdensity}). Values considered for the parameters $\lambda_\Phi$, $\lambda_{\Phi H}$, $\xi_H$ and $\xi_\Phi$ are motivated by inflation and preheating (see Sec.~\ref{eq:paramspace}).

\section{Cosmological history}
\label{sec:sec3}

As mentioned above, we study, for concreteness and simplicity, a generalised dark-sector Higgs inflation, inspired by e.g. Ref.~\cite{Bezrukov:2007ep}, where the inflaton $\eta$ is well-approximated by $\sqrt{2}|\Phi|$ in our model.  By postulating an explicit inflation scenario, the relic density may be connected to the primordial power spectra, including gravitational waves, and the reheating temperature; the details of reheating being model-dependent.\footnote{A $Z'$ reheaton may still be consistent with inflaton potentials other than what we assume here.}  

\subsection{Inflationary epoch} 
\label{sec:sec3A}

\begin{figure}[t!]
\centering
\includegraphics[width=0.49\textwidth]{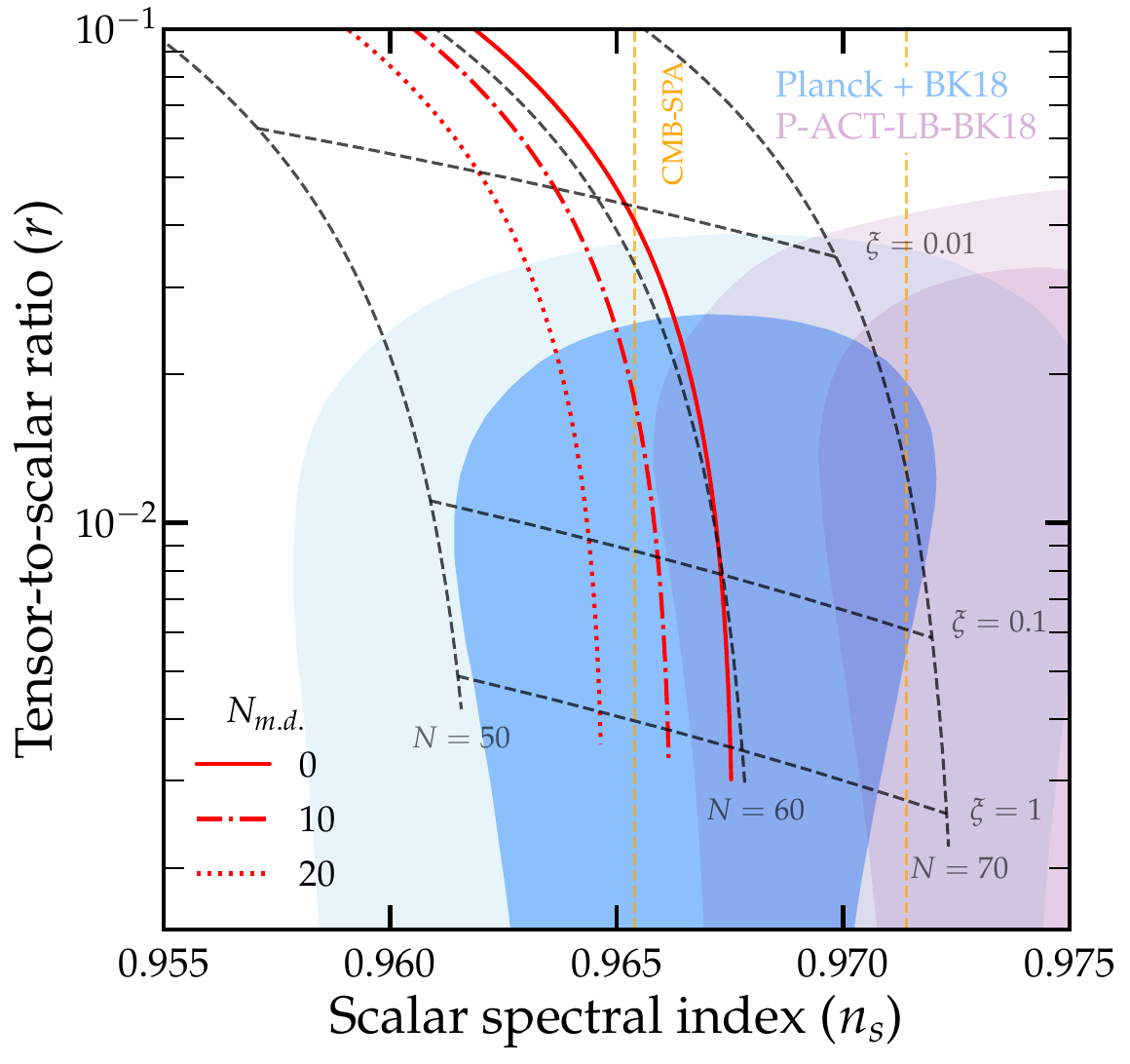} \includegraphics[width=0.49\textwidth]{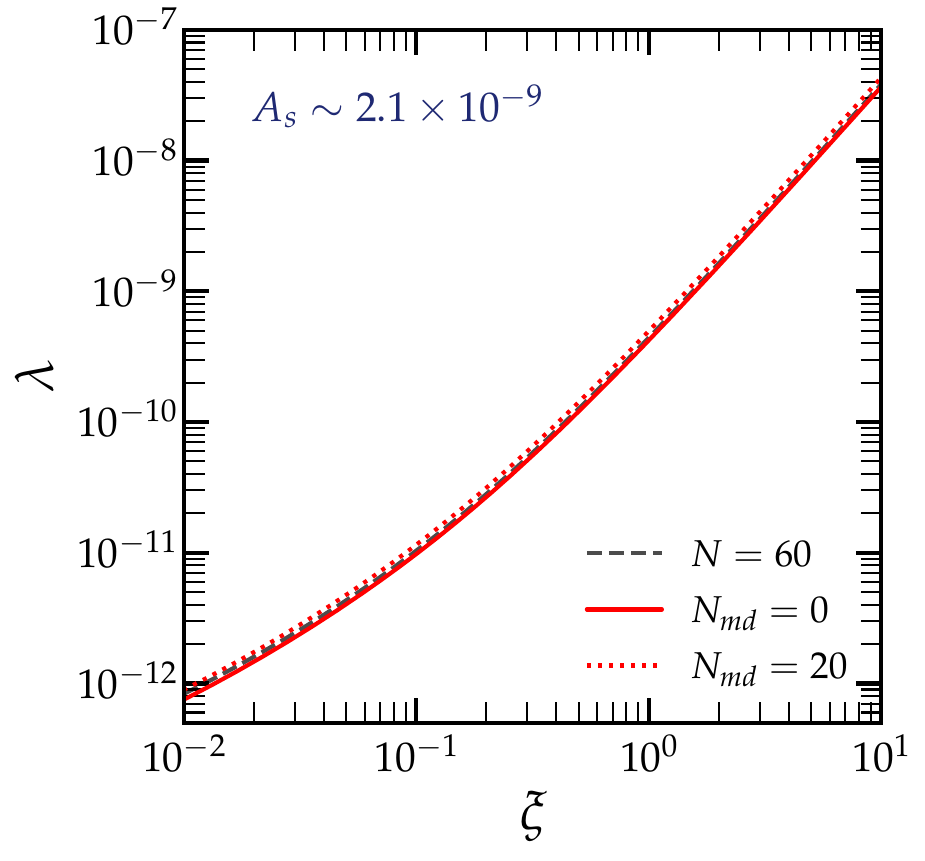}
\caption{
\label{fig:rnsplot}
\textit{Left:} we summarise the inflationary predictions for the generalised model of Higgs inflation in the $r$-$n_s$ plane. We use dashed black lines to illustrate the dependence on the e-folds of inflationary expansion ($N$)  before the Hubble exit of the pivot scale ($k_*$) and the effective non-minimal coupling of the inflaton ($\xi$). 
We draw some contours in red to exhibit the dependence on the e-folds of primordial matter-domination, which is generally non-zero in our setup. The light and dark shaded regions correspond to 95\% and 68\% confidence intervals at the pivot scale $k_* = 0.05\ \text{MPc}^{-1}$; the model is in excellent agreement with the combined Planck and BK18 fit~\cite{Planck:2018jri,BICEP:2021xfz}, as well as SPT~\cite{SPT-3G:2025bzu} and ACT~\cite{ACT:2025fju}, but there is a small tension with the inclusion of DESI~\cite{DESI:2025zgx} data by Ref.~\cite{ACT:2025fju}. \textit{Right:} the normalisation of the scalar power spectrum is achieved by the parameter hierarchy $\lambda \ll \xi$, with the exact functional dependence plotted numerically above. 
The red contours show that this depends negligibly on $N_{m.d.}$. 
}
\end{figure}

A non-minimal gravitational coupling of a scalar inflaton field $\mathcal{L} \supset \xi\eta^2R$ with quartic potential $\frac{\lambda}{4}\eta^4$ represents a well-motivated inflationary model~\cite{Bezrukov:2007ep}; it provides an adequate fit to a combination of \textit{Planck} 2018~\cite{Planck:2018jri}, \textit{Bicep/Keck} (BK18~\cite{BICEP:2021xfz}), \textit{Atacama Cosmology Telescope} (ACT-DR6~\cite{ACT:2025tim}) and \textit{South Pole Telescope} (CMB-SPA~\cite{SPT-3G:2025bzu}) data,\footnote{At present, there is an interesting tension~\cite{ACT:2025fju} once BAO data from the \textit{DESI} collaboration~\cite{DESI:2025zgx} are also included (see also Ref.~\cite{SPT-3G:2025bzu}). There is some debate over the status of these constraints~\cite{Kallosh:2025rni,Ferreira:2025lrd, SPT-3G:2025bzu}. Nonetheless, one could consider a different inflationary completion for our setup with a slightly higher scalar spectral index.} for the case where $\xi \gtrsim \mathcal{O}(0.01)$  and the large field regime $\xi\eta^2 \gtrsim M_P^2$ is satisfied primordially by the homogeneous part $\eta(t)$. Predictions for the tensor-to-scalar ratio, $r = A_t/A_s$, and the scalar spectral index, $n_s$, associated to the scalar, $\Delta^2_{\mathcal{R}} \simeq A_s\left( k/k_*\right)^{n_s - 1}$, and tensor, $\Delta^2_{t} \sim A_t$, power spectra, at a reference scale $k_*$,
are recapitulated in Figure \ref{fig:rnsplot}; left panel. In particular, we illustrate their dependence on $\xi$ and $N$, the number of e-folds of inflationary expansion required to produce the CMB pivot scale $k_*$
\begin{equation}
\begin{split}
    N(k_*) &\equiv \int^{t_*}_{t_{end}} H[\eta(t)]\ \mathrm{d}t \\ &\simeq  \Delta N + \frac{1}{4}\log \frac{\rho_{\text{eq}}}{\rho_{\text{reh}}} - \log \frac{k_*}{a_0 H_0} 
    + \log \frac{H_{k_*}}{H_{\text{eq}}}
    + \log( 219h\Omega_0),
\end{split}
\end{equation}
where $t_*$ indicates the cosmic time when the scale $k_*$ exited the horizon and $t_{\text{end}} > t_*$ is the time when the inflationary era ends. 

As the second equality shows, the e-folds are fixed by the post-inflationary expansion (see Section \ref{sec:sec3Ba}) approximating instantaneous transitions between epochs~\cite{Liddle:2003as}, which can then be related to $r$ and $n_s$ predictions through the corresponding background inflaton value $\eta_*$, whose slow-roll dynamics power the expansion rate $H = \frac{\dot{a}}{a}$. Here, subscripts `$k_*$', `reh', `eq' and `$0$' respectively denote values at the moment of the horizon exit of the pivot scale, the end of reheating, matter-radiation equality and the present epoch; we denote by $\Delta N$ the number of e-folds of reheating. 

The reheating epoch will initially mimic radiation domination, as we explain in Section \ref{sec:sec3Bb}. However, the thermalisation time-scale will in general be longer than the time-scale for the expansion rate to render the $Z'$ bath non-relativistic, resulting in a primordial matter-like epoch (let `m.d.' refer to when it begins) that modifies our predictions, see the red contours in the left panel of Figure \ref{fig:rnsplot}. Hence,
\begin{equation}
    \Delta N \simeq \frac{1}{4}\log \frac{\rho_{\text{end}}}{\rho_{\text{m.d.}}} + \frac{1}{3}\log \frac{\rho_{\text{m.d}}}{\rho_{\text{reh}}}.
\end{equation}
The scalar spectral amplitude, $A_s \simeq 2.1 \times 10^{-9}$ ~\cite{Planck:2018vyg}, adds a further constraint at the inflationary scale (see Figure \ref{fig:rnsplot}). For the purposes of this paper, we focus on the parameter space $\xi \lesssim 1$, because this is certainly free from possible unitarity issues identified and discussed in Refs.~\cite{Barbon:2009ya,Burgess:2009ea,Bezrukov:2010jz,Burgess:2014lza,Bezrukov:2014ipa}. The inflaton self-interaction $\lambda \ll 1$ is consequently very suppressed.

It is possible to generalise this single-field treatment to our multi-scalar setting as follows. Due to Hubble friction on the angular field velocities, the slow-roll inflationary trajectory of the displaced scalar background fields, $\eta(t)$, is stabilised along a straight radial trajectory in the ($|\Phi|, h$) field-space~\cite{Kaiser:2013sna}. We are interested in the scenario where the primordial energy density is stored in the dark sector scalar, that is, where this trajectory is at most a small-angle displacement from the $|\Phi|$ axis (so that $\eta \simeq \sqrt{2}|\Phi|$). For $\xi_H = 0$, there is no Higgs component for $\lambda_{\Phi H} >0$, while for $\lambda_{\Phi H} < 0$ and $|\lambda_{\Phi H}| \ll \lambda_H$ the Higgs component is suppressed with a misalignment angle $\sin \psi \simeq \sqrt{\frac{- \lambda_{\Phi H}}{\lambda_{H}}}$~\cite{Ballesteros:2016xej}. In the more general parameter space of $\xi_H \sim \xi_\Phi$ (but not $\xi_H \gg \xi_\Phi$), there is again a suppressed Higgs component with $\sin \psi \sim  \sqrt{\frac{\lambda_\Phi}{\lambda_{H}}}$ for $|\lambda_{\Phi H}| \ll \lambda_\Phi$~\cite{Sopov:2022bog}.
When there is a misalignment we can still identify, up to subleading $\mathcal{O}(\psi)$ corrections, the ``effective'' inflaton couplings as $\lambda \simeq \lambda_\Phi$ and $\xi \simeq \xi_{\Phi}$. Then, in each of these cases, and to the level of precision required, our subsequent results are the same provided $|\lambda_{\Phi H}| \ll \lambda_\Phi$. This condition ensures that SM thermalisation is sufficiently hampered, and hence, the reheating temperature consequently low, to avoid the standard scenarios for DM production in the $U(1)_D$ model.\footnote{A small Higgs component in the inflationary background will produce an initially high-temperature, but subdominant and dilute, radiation bath. Any associated thermally produced $\chi$ and $Z'$ would then be washed out by red-shifting and non-thermal particle production at later times. }

The smallness of the self-coupling $\lambda_\Phi$, required by $\xi_\Phi \lesssim 1$, 
also motivates restrictions on the sizes of $g_D$ and $\lambda_{\Phi H}$. 
The 1-loop beta function for the parameter $\lambda_\Phi$ is given by
\begin{equation}
    \beta_{\lambda_\Phi} = \frac{1}{16\pi^2} \left[ 5 \lambda_{\Phi}^{2} + 2 \lambda_{\Phi H}^{2} - 6 g_{D}^{2} \lambda_{\Phi} + 6 g_{D}^{4}\right].
\end{equation}
While the radiative correction is strictly positive, so that there is no stability issue in the inflaton direction, the gauge-coupling and scalar interactions may be seen to provoke a fine-tuning issue if $g_D^2,\lambda^2_{\Phi H} \gg \lambda_{\Phi}$, because it would then follow that $|\beta_{\lambda_\Phi}| \gg \lambda_{\Phi}$.

In summary, then, we focus on the following parameter space, motivated by our assumptions about the background dynamics of the inflationary era (see again the right panel of Figure 
\ref{fig:rnsplot} for the exact relation between $\lambda_\Phi$ and $\xi_\Phi$)
\begin{equation}\label{eq:paramspace}
    \xi_H \lesssim \xi_\Phi \lesssim 1,\qquad \lambda_{\Phi H}, g_D^2 \ll \lambda_{\Phi} \lesssim 4.2 \times 10^{-10},\qquad v_D < 10^{13}\ \text{GeV}.
\end{equation}
The smallness of the $g_D$ coupling is consequential for the primordial production of $Z'$, while the smallness of $\lambda_{\Phi\,H}$ ensures the SM radiation bath originates from $Z^\prime$ decay, which acts as a reheaton. For completeness, we have considered a conservative upper bound on $v_D$ so that we can ignore dimensionful couplings during inflation and preheating (Section \ref{sec:sec3Ba}); and also satisfy bounds on the relic gravitational waves originating from the near-global string network (see Section \ref{sec:sec4}).

\paragraph{Inflationary $Z'$ production}

The non-perturbative production of gauge bosons during inflation has been studied in a variety of contexts. In models where the gauge kinetic term is canonical (i.e. conformally-invariant), production generally takes place through the longitudinal mode, so a sub-Hubble mass must be included, for example, using a St\"uckelberg mechanism (see e.g. Refs.~\cite{Graham:2015rva,Ozsoy:2023gnl}). The alternative, which we consider here, is to use a Higgs mechanism. The case of the charged inflaton~\cite{Lozanov:2016pac,Ema:2016dny,Firouzjahi_2021} (or displaced spectator~\cite{Dror:2018pdh,Sato:2022jya}) is particularly interesting, since it creates a mass-splitting with the transverse modes, and introduces a time-dependent mass subject to the dynamics of the background field. Nonetheless, in the slow-roll epoch, we will see that these additional effects are rather mitigated and the initial spectrum resembles the super-horizon spectrum in the St\"uckelberg case, with the differences being much more dramatic during reheating (see Section~\ref{sec:sec3B}). 

\begin{figure}[t!]
\centering
\includegraphics[width=0.49\textwidth]{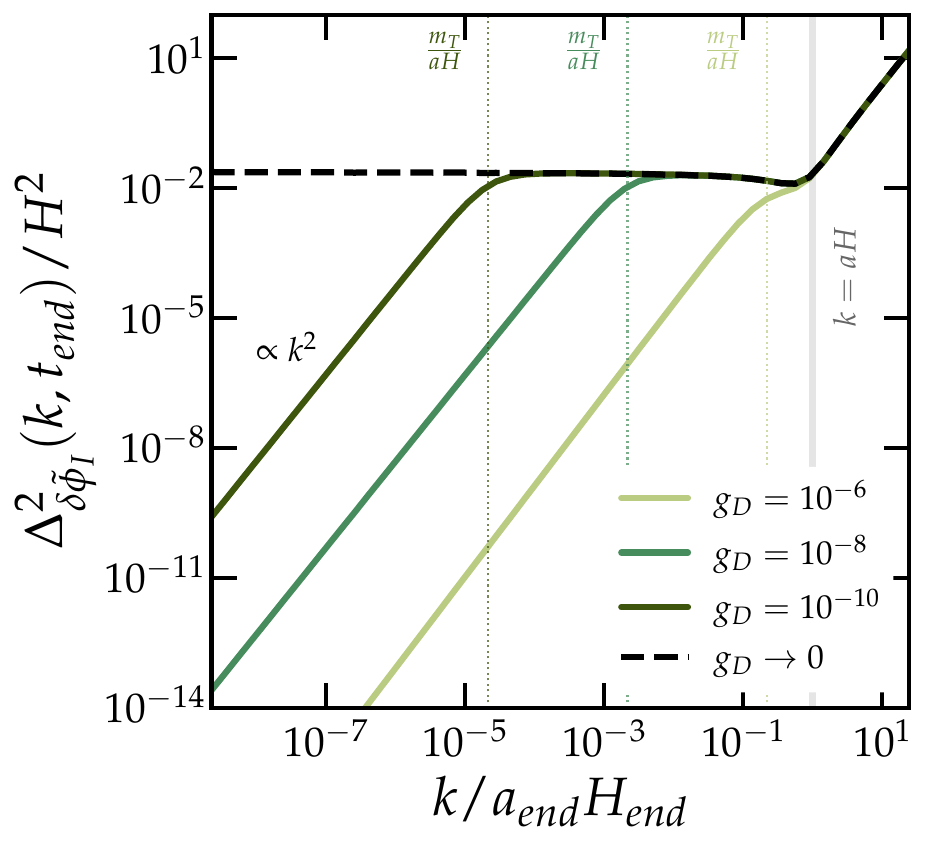}
\includegraphics[width=0.49\textwidth]{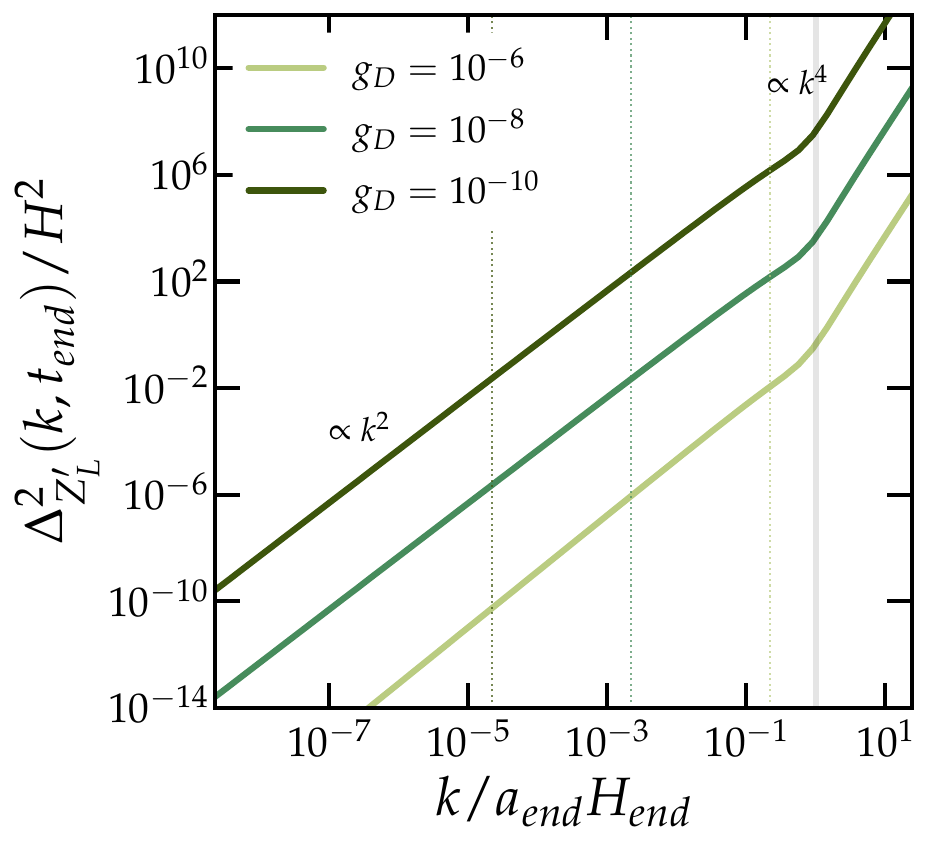}
\caption{
\label{fig:powerspectra} Power spectra at the end of inflation for the rescaled Goldstone modes in Coulomb gauge $\delta\tilde{\phi}_{I\mathbf{k}}$ (\textit{left}) and the corresponding longitudinal polarisation modes of the $Z'$ in unitary gauge (\textit{right}). The vectorial $\propto k^2$ contribution to the energy density per logarithmic interval is manifest in both cases, resulting in  dramatic isocurvature suppression in the long wavelength limit~\cite{Graham:2015rva,Ozsoy:2023gnl}. 
}
\end{figure}

In this subsection, we analyse the case where the charged inflaton is non-minimally coupled (as in Ref.~\cite{Ema:2016dny}), but where the near-global limit of the gauge theory is assumed ($g_D^2 \ll \lambda_\Phi$). Our result is summarised in Figure \ref{fig:powerspectra}, explained further below, and comprises the initial conditions for the longitudinal modes in the post-inflationary era.  

Following Ref.~\cite{Lozanov:2016pac}, we introduce a field parametrisation 
\begin{equation}
\begin{split}
    \Phi(t,\mathbf{x}) &= \frac{1}{\sqrt{2}} \rho(t,\mathbf{x}) e^{ig_D \theta(t,\mathbf{x})},\qquad \quad \ \ \text{with} \quad \langle \rho(t,\mathbf{x}) \rangle = \varrho(t) \neq 0; \\
    G_\mu(t,\mathbf{x}) &= Z'_\mu(t,\mathbf{x}) + \partial_\mu \theta(t,\mathbf{x}),\qquad \quad \  \text{with} \quad \langle G_\mu \rangle = 0; \\
    &= [ G_0 (t,\mathbf{x}),\ \partial_i G^{\parallel}(t,\mathbf{x}) + G_i^\perp(t,\mathbf{x}) ] ,\qquad \quad  \text{with} \  \ \quad \partial_iG^{\perp i} = 0,\ G^L_i \equiv \partial_i G^\parallel;
\end{split}
\end{equation}
invariant under the gauge transformation $\rho \rightarrow \rho$, $\theta \rightarrow \theta - \alpha$. We may then define longitudinal ``L'' and transverse ``T'' Fourier components
\begin{equation}
    \mathbf{G}^L_\mathbf{k}= \mathbf{\epsilon}_\mathbf{k}^L G_\mathbf{k}^L \qquad  \mathbf{G}^\perp_\mathbf{k}= \sum_{\lambda = \pm} \mathbf{\epsilon}_\mathbf{k}^{T\pm} G_\mathbf{k}^{T\lambda},
\end{equation}
using polarisation vectors satisfying 
\begin{equation}
    \begin{split}
    &\mathbf{\epsilon}_\mathbf{k}^L = \mathbf{\epsilon}^{L*}_{-\mathbf{k}},\quad \quad \mathbf{\epsilon}_\mathbf{k}^{L*} \cdot \mathbf{\epsilon}^{L}_{\mathbf{k}} = 1,\quad \qquad i\mathbf{k}\cdot \mathbf{\epsilon}^{L}_{\mathbf{k}} = k,\quad  \ \ \ i\mathbf{k} \times \mathbf{\epsilon}^{L}_{\mathbf{k}} = 0,\\
    &\mathbf{\epsilon}_\mathbf{k}^{T\pm} = \mathbf{\epsilon}^{T\pm*}_{-\mathbf{k}},\quad \mathbf{\epsilon}_\mathbf{k}^{T\lambda*} \cdot \mathbf{\epsilon}^{T\lambda'}_{\mathbf{k}} = \delta^{\lambda \lambda'},\quad i\mathbf{k}\cdot \mathbf{\epsilon}^{T\pm}_{\mathbf{k}} = 0,\quad i\mathbf{k} \times \mathbf{\epsilon}^{T\pm}_{\mathbf{k}} = \pm k \mathbf{\epsilon}^{T\pm}.
    \end{split}
\end{equation}
To fix a gauge for our dark sector, we then work in a Cartesian parametrisation for $\Phi$,
\begin{equation}
    \Phi(t,\mathbf{x}) = \frac{\varphi(t) +\delta\phi_R(t,\mathbf{x}) + i\delta\phi_I(t,\mathbf{x})}{\sqrt{2}},
\end{equation}
using the global rephasing symmetry to set $\langle \text{Im}\ \Phi\rangle = 0$ so that $|\varphi(t)| = |\langle \text{Re}\ \Phi\rangle| = \varrho(t)$. Then, as it is most convenient to study the near global limit ($g_D^2 \ll \lambda_\Phi$), and to avoid the field coordinate singularity when $\varrho(t) = 0$ during the post-inflationary dynamics, we will work in the Coulomb gauge such that
\begin{equation}\label{eq:gaugemodes}
    \delta \phi_{R\mathbf{k}} = \delta \rho_\mathbf{k},\quad \delta \phi_{I\mathbf{k}} = - \frac{ g_D \varrho}{k} G_\mathbf{k}^L,\quad  {Z'}_\mathbf{k}^{T\pm} = G_\mathbf{k}^{T\pm}\quad\text{and} \quad {Z'}_\mathbf{k}^L = 0.
\end{equation}
In unitary gauge, one has instead that $\delta \phi_{I\mathbf{k}} = 0$ and ${Z'}_\mathbf{k}^L =G_\mathbf{k}^L $. In the parameter space (\ref{eq:paramspace}), not only the inflationary production, but the post-inflationary enhancement of the transverse components will be suppressed.\footnote{The modes of temporal component of the vector decouple and are non-dynamical~\cite{Graham:2015rva}; they may be expressed in the Coulomb gauge, via a linearised constraint equation~\cite{Lozanov:2016pac}, in terms of the other modes
\begin{equation}
    X_{0\mathbf{k}} = \frac{g_D}{2} \frac{[\varphi'\delta\phi_{I\mathbf{k}} - \varphi\delta\phi_{I\mathbf{k}}']}{(k/a)^2 + (g_D\varphi/2)^2}.
\end{equation}} We focus instead on the much richer dynamics of the Goldstone modes $\delta\phi_{I\mathbf{k}}$. 

If we introduce a conformal rescaling which canonically normalises the Fourier mode kinetic terms
\begin{equation}\label{eq:moderescaling}
    \delta \tilde \phi_{I\mathbf{k}} = \frac{ak}{\sqrt{k^2 + g_D^2\tilde\varphi^2}}  \delta \phi_{I\mathbf{k}} \qquad \text{with} \qquad \tilde \varphi = a\varphi \qquad \text{and} \qquad k = |\mathbf{k}|
\end{equation}
then, following Ref.~\cite{Ema:2016dny}, the conformal-time quadratic action for the $\delta \tilde\phi_{I\mathbf{k}}$ is 
\begin{equation}\label{eq:quadaction}
    S_{\phi_I}^{(2)} = \frac{1}{2} \int \frac{\mathrm{d}^3\mathbf{k}\mathrm{d}\tau}{(2\pi)^3} \left[ |\delta \tilde\phi'_{I\mathbf{k}}|^2 - (k^2 + m^2_{I,\text{eff}})|\delta \tilde\phi_{I\mathbf{k}}|^2 \right],
\end{equation}
where
\begin{equation}\label{eq:effmass}
    m^2_{I,\text{eff}} = m^2_{T,\text{eff}}  - \frac{k^2}{k^2+m_{T,\text{eff}}^2}\left[ \frac{m_{T,\text{eff}}''}{m_{T,\text{eff}}} - \frac{3(m_{T,\text{eff}}')^2}{k^2+m_{T,\text{eff}}^2} \right] \qquad \text{with} \qquad m_{T,\text{eff}} = g_D\tilde\varphi,
\end{equation}
We can then use the background equation of motion
\begin{equation}\label{eq:bgeom}
    \tilde \varphi '' + \underbrace{\left[\lambda_\Phi(\tilde \varphi^2-a^2v_D^2) + a^2\left(\xi_\Phi + \frac{1}{6}\right)R\right]}_{-m_{T,\text{eff}}''/m_{T,\text{eff}}}\tilde\varphi = 0 \qquad \text{with} \qquad R = \frac{-6a''}{a^3}
\end{equation}
to exhibit the dependence on $\lambda_\Phi$ and $\xi_\Phi$ in the effective mass of the Goldstone modes, which generally differs from the effective mass of the transversely polarised modes ($m_{T,\text{eff}}$) due to the displacement of the background field away from the global minimum. 

Let us define the quantity $q=g_D^2/\lambda_{\Phi} \ll 1$ and ignore dimensionful couplings. Note that for the ultraviolet regime of wavenumbers $k \gg m_{T,\text{eff}}, \sqrt{3q\tilde\varphi'}$ it follows that
\begin{equation}
    m^2_{I,\text{eff}} \simeq \lambda_\Phi(1 + q)\tilde \varphi^2 + a^2\left(\xi_\Phi + \frac{1}{6}\right)R. 
\end{equation}
Hence, it is possible to treat the rescaled Goldstone modes as those of a non-minimally coupled scalar-singlet. However, for finite $g_D$, this cannot be the case in the deep infrared. For example, for $k \ll \sqrt{q}m_{T,\text{eff}}$ the Goldstone modes behave as transverse components with mass $m_{T,\text{eff}}$ (up to corrections from the non-minimal coupling that vanish for $k\rightarrow 0)$; while in intermediate regimes, there is also a non-trivial dependence on the last term in (\ref{eq:effmass}). 

Following the standard procedure, we canonically quantise the theory in the Heisenberg picture. In particular, we may expand the Goldstone field operator as $\widehat{\delta \tilde\phi}_{I\mathbf{k}}(\tau) = \chi_{k}(\tau)\hat a_{\mathbf{k}} + \chi^*_{k}(\tau)\hat a^\dagger_{-\mathbf{k}}$ and require $\chi_{k} \chi^{\prime*}_{k}  - \chi_{k}' \chi^*_{k} = i$. The power spectrum for $\delta\phi_I$ is then constructed from the coincident 2-point function of the rescaled operator
\begin{equation}
    \langle \delta\tilde\phi^2_I\rangle=\lim_{\mathbf{x}' \rightarrow \mathbf{x}} \left\langle \widehat{\delta \tilde\phi}_{I}(\tau,\mathbf{x'})\widehat{\delta \tilde\phi}_{I}(\tau,\mathbf{x})  \right\rangle = \int \frac{dk}{k} \frac{k^3}{2\pi^2}|\chi_k(\tau)|^2 = \int \frac{dk}{k} \Delta^2_{\delta\tilde\phi_I}(k,\tau) 
\end{equation}
so that, using (\ref{eq:moderescaling})
\begin{equation}\label{eq:powspecgoldstone}
     \Delta^2_{\delta\phi_I}(k,\tau) = \frac{k^2+g_D^2\tilde\varphi^2}{a^2k^2}\Delta^2_{\delta\tilde\phi_I}(k,\tau)  \qquad \text{and} \qquad \Delta^2_{G^L_\mathbf{k}} = \frac{a^2k^2}{g_D^2\tilde\varphi^2}\Delta^2_{\delta\phi_I}(k,\tau).
\end{equation}

Note that the time dependence of the field operator is encoded in the mode functions, which satisfy the linearised equations of motion given by (\ref{eq:quadaction})
\begin{equation}\label{eq:modeeom}
    \delta \chi''_{k} + \underbrace{(k^2 + m^2_{I,\text{eff}})}_{\omega^2_{Ik}}\delta \chi_{k} = 0.
\end{equation}
The inflaton drives a quasi-de Sitter expansion $\tau \simeq \frac{-1}{aH}$ with $H$ approximately constant. At sufficiently early times ($a\rightarrow 0$), each mode may be considered to be both sub-horizon ($k \gg aH$) and ultra-relativistic ($k \gg m_{I,\text{eff}}$, $\sqrt{\lambda_\Phi}\tilde\varphi$). We then assume the initial condition to be the WKB solution associated to the Bunch-Davies vacuum state
\begin{equation}\label{eq:bunchdavies}
     \chi_{k}(\tau) \longrightarrow \frac{e^{-ik\tau}}{\sqrt{2k}}.
\end{equation}
The general picture is that inflation maps the relativistic, small-scale, sub-horizon modes into non-relativistic, large-scale, super-horizon modes under time evolution with (\ref{eq:modeeom}). For $m_{I,\text{eff}} < H$, the latter modes may be treated as a classical ensemble of many waves with a Gaußian-distributed amplitude with power spectrum (\ref{eq:powspecgoldstone}). 

We obtain the full power spectrum at the end of inflation by numerically solving the mode equations (\ref{eq:modeeom}) in cosmic time. This is plotted in Figure \ref{fig:powerspectra} for different values of $g_D$, and compared with the imaginary part of $\Phi$ for a global $U(1)_{D}$, i.e. $g_D \rightarrow 0$, where the good agreement for $k \gg m_{T,\text{eff}}$ is manifest. We also compare this to the unitary gauge power spectrum for the longitudinal modes of the $Z'$ (both agree at $k\rightarrow 0$), obtained by the rescaling in (\ref{eq:gaugemodes}). In the super-horizon regime, we then recapitulate the result of Ref.~\cite{Graham:2015rva} for the case of the charged non-minimally coupled inflaton, viz.
\begin{equation}
    \Delta^2_{G^L_\mathbf{k}}(k, t_{\text{end}}) \simeq  \left( \frac{k H_{k}}{2\pi m_{T,\text{eff}}(t_\text{end})} \right)^2\ \quad \text{when}\quad k < a_\text{end} H_\text{end}.
\end{equation}

Neglecting interactions, each rescaled mode contributes to the physical energy density stored in the Goldstone for the $Z'$ as
\begin{equation}
    \langle \rho_{G} \rangle = \frac{1}{2a^4}\int \frac{\mathrm d^3 \mathbf k}{(2\pi)^3} \left[ |\chi'_k|^2 + (k^2+m^2_{I,\text{eff}})|\chi_k|^2\right]
\end{equation}
Hence, we see that the characteristic $\propto k^2$ inflationary spectrum for superhorizon, non-relativistic and sub-Hubble mass vector modes~\cite{Graham:2015rva,Ozsoy:2023gnl} renders the isocurvature perturbation in $Z'$ completely negligible in the long-wavelength limit relevant to CMB constraints.

\subsection{Post-inflationary reheating}
\label{sec:sec3B}
As the $Z'$ is Higgsed by the inflaton, production can continue even after the inflationary era, and is particularly efficient due to non-perturbative effects.  We show that, for a sufficiently long reheating epoch, and sufficiently weak portal couplings, the post-inflationary dynamics inexorably leads to a bath of $Z'$ dominating the energy budget of the Universe; the $Z'$ ultimately being a lighter state than the inflaton (for $g_D^2 \ll \lambda_{\Phi}$). As the linear and perturbative approximations for the interactions of the scalar components quickly breakdown in the early stages of reheating, we study this process using lattice simulations. This not only improves calculational accuracy, but also accounts for the relic gravitational waves radiated by the large field gradients, including local strings, developed in this epoch (see Section \ref{sec:sec4}). We then study the perturbative freeze-in production of dark matter ($\chi$) from the non-thermal $Z^\prime$ reheatons, before they produce the SM bath. 

\subsubsection*{Non-perturbative reheating era}
\label{sec:sec3Ba}

In Section \ref{sec:sec3A}, we have motivated an explicit model for the inflationary epoch. For $\xi_\Phi \lesssim 1$, the small field-regime ($\varphi < M_P/\sqrt{\xi_\Phi}$) is always satisfied in the post-inflationary era, and we can neglect the non-minimal couplings in what follows.\footnote{It is worth reiterating that we chose this limit simply to avoid possible unitarity issues associated with large $\xi$. In fact, a large non-minimal coupling should simply expedite the onset of longitudinal $Z'$ domination through the development of ``spikes'' during preheating (see Refs.~\cite{Ema:2016dny,Sfakianakis:2018lzf,DeCross:2016cbs, DeCross:2016fdz}), but results in additional matter-domination as the inflaton would oscillate in an initially quadratic minimum. } 
Neglecting curvature, the inflaton zero-mode then undergoes coherent oscillations in an approximately quartic potential. In particular, the solution to (\ref{eq:bgeom}) is a Jacobi elliptic function of constant amplitude,
so that $\varphi$ has a red-shifting envelope $\propto a^{-1}$~\cite{Greene_1997}, producing a radiation dominated expansion $a \propto \sqrt{t}$ and inducing mass contributions $\propto a^{-1}$ for the fields coupled at tree-level to $\Phi$. 

\paragraph{Preheating and lattice simulations} 

In the presence of the oscillatory background solution $\tilde\varphi(\tau)$, the mode functions for the conformally rescaled inflaton ($\phi_k$) and the Goldstone ($\chi_k$), excited during inflation but initially energetically sub-dominant, satisfy 
\begin{equation}
    \phi''_{k}(\tau) + \underbrace{[k^2 + 3\lambda_{\Phi}\tilde\varphi^2(\tau)]}_{\omega^2_{k\phi}} \phi_{k}(\tau) = 0\quad  \text{and}\quad \chi''_{k}(\tau) + \underbrace{[k^2 + \  \lambda_{\Phi}\tilde\varphi^2(\tau)]}_{\omega^2_{k\chi}} \chi_{k}(\tau) = 0,
\label{eq:modeeqs}
\end{equation}
for $k \gg  m_{T,\text{eff}}, \sqrt{3q\tilde\varphi'}$, assuming a linear approximation. 
Note that, in the higher frequency regime, the modes evolve like those of a complex scalar inflaton (see e.g. Refs.~\cite{Greene_1997,Tkachev:1998dc}), a fact which is intuitively explained by the Goldstone Boson Equivalence Theorem (GBET) and the near-global limit of the gauge symmetry. 

It is well-known that these equations have exponentially growing solutions $\propto \exp(\mu^I_k \tau)$  where $\mu^I_k$ is a Floquet index, $I 
\in \{\phi,\chi\}$ and $\text{Re}\{\mu_k\} >0$ for specific ranges of $\kappa \equiv \frac{k}{\sqrt{\lambda}\tilde\varphi_{\text{end}}} $~\cite{Greene_1997}.\footnote{Due to $|\lambda_{\Phi H}|,g_D^2 \ll \lambda_\Phi$, fluctuations in the Higgs, and the transverse components of the Abelian gauge boson are not enhanced and may be studied perturbatively.}  In particular, the $\phi$ modes are excited primarily in the ``instability band'' $\frac{3}{2}< \kappa^2 < \sqrt{3}$, with maximal growth $\mu^\phi(\kappa_\text{max} \simeq 1.27) \simeq 0.0359$; while $\chi$ modes are the more strongly excited, in a band $ \kappa^2 < \frac{1}{2}$, with $\mu^\chi(\kappa_\text{max} \simeq 0.47) \simeq 0.147$. Of course, in the latter case, the parametric resonance is disrupted at small scales when the Goldstones behave vectorially. Note that the horizon size at the end of inflation is e.g. $ \kappa^2_{\text{hor},\text{end}} \simeq 0.44$ for $\xi_\Phi \simeq 0.5$, meaning that the modes of interest primarily lie in (or quickly enter) the sub-horizon regime. If the dimensionful coupling were sufficiently large, the inflaton zero-mode in principle crosses a convex region of the potential during its oscillations, leading to tachyonic masses $m^2_\tau < 0$ for the fluctuations. For simplicity, we work in a parameter space where this is absent, requiring (conservatively) $v_D < H_{\text{inf}} \sim 10^{13}$ GeV.\footnote{This bound is also conservative enough to ensure that, if a scaling string network is produced, it is not excluded by pulsar timing arrays (see Section \ref{sec:sec4}). }  

\begin{figure}
\centering 
\includegraphics[width=0.45\textwidth]{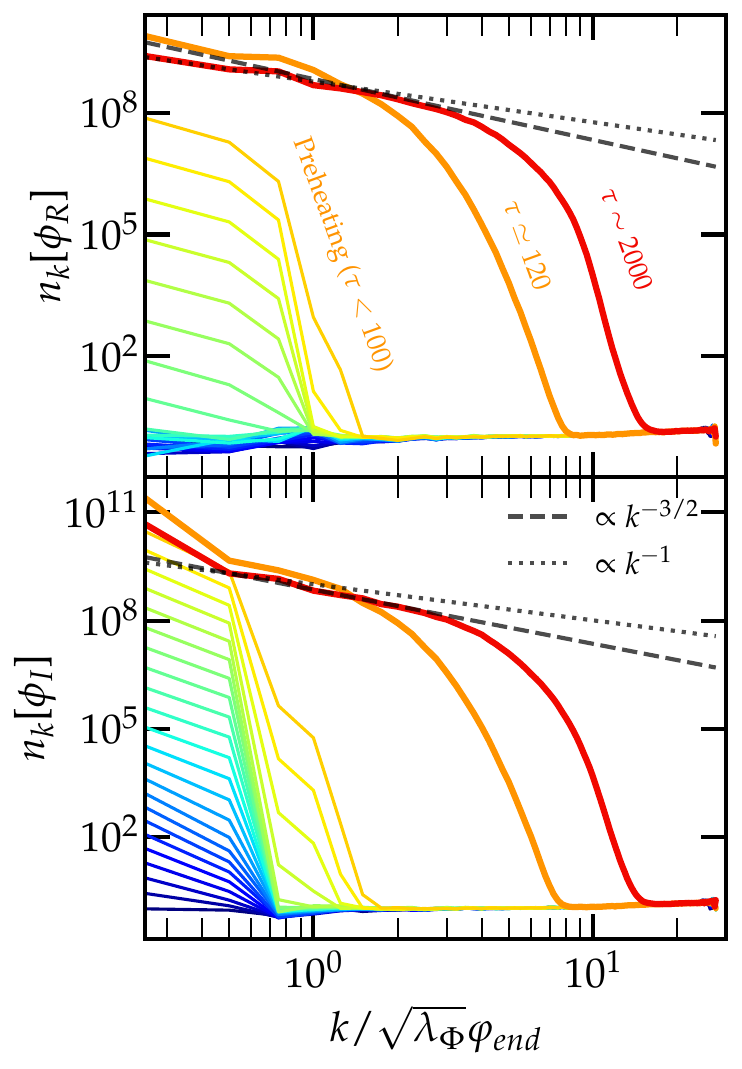}
\includegraphics[width=0.45\textwidth]{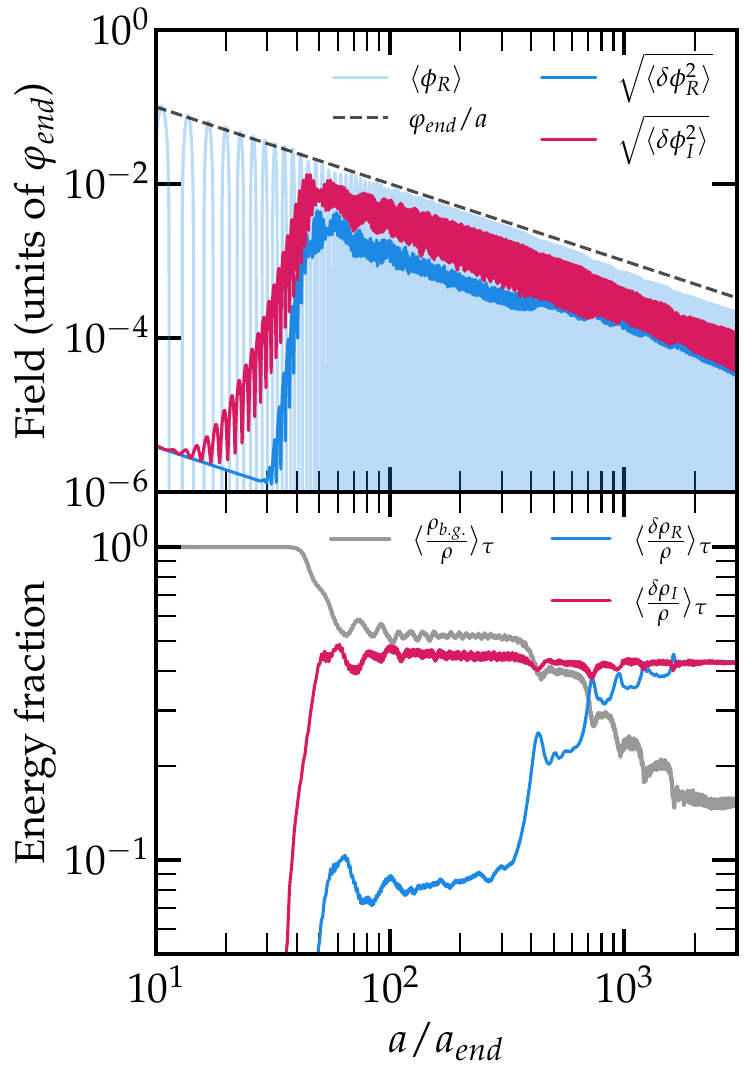}
\caption{\label{fig:lattice}\textit{Left:} the occupation number $n_k = \rho_k/\omega_k$ spectra for the real and imaginary components of the $\Phi$ field, with bluer contours for $t \sim t_{\text{end}}$ becoming redder at later times. See text for details. \textit{Right upper:} the oscillating field zero mode compared with the root-mean-squared of the field fluctuations, which are amplified by preheating, restoring the potential minimum to the origin. 
\textit{Right lower:} we compare estimates for the energy fraction in the background ($\rho_{bg}$) as well as fluctuations in the inflatons and Goldstones using $\delta\rho_X \sim(\dot{\delta\phi_X})^2$. }
\end{figure}

This initial explosive phase of non-perturbative particle production after inflation (\textit{preheating}) is brief. Within a few oscillations, the linearised approximation breaks down and the parametric resonance is terminated by back-reaction effects. 
Lattice numerical simulations of the inhomogeneous field equations, which we performed using the $\mathcal{C}$osmo$\mathcal{L}$attice~\cite{Figueroa:2020rrl,Figueroa:2021yhd} package, then become essential to model the post-inflationary dynamics accurately.
The evolution of the sub-horizon power spectra for the inflaton and Goldstone modes during the preheating epoch; drawn from a typical simulation with benchmark values: $\lambda_\Phi = 2.5 \times 10^{-10}$ ($\xi_\Phi = 0.5$), $ \lambda_{H\Phi} = 10^{-11}$, $\lambda_H = 0.25$, $\varphi_{\text{end}} = 1.33\,M_{P}$, infrared wavenumber cutoff $k_{\text{IR}} = 0.25\sqrt{\lambda_\Phi}\varphi_{\text{end}}$
, number of lattice sites per dimension $N = 128$, and time-step $d \tau = 10^{-3}\sqrt{\lambda_\Phi}\varphi_{\text{end}}$ (sufficiently small to ensure energy conservation over the integration time); are illustrated in Figure \ref{fig:lattice}. 
(We also include the SM Higgs in the simulation as a proxy for the SM radiation energy density with which it rapidly thermalises.) 

\paragraph{Non-linear evolution} 
\begin{figure}
\centering
\includegraphics[width=0.48\textwidth]{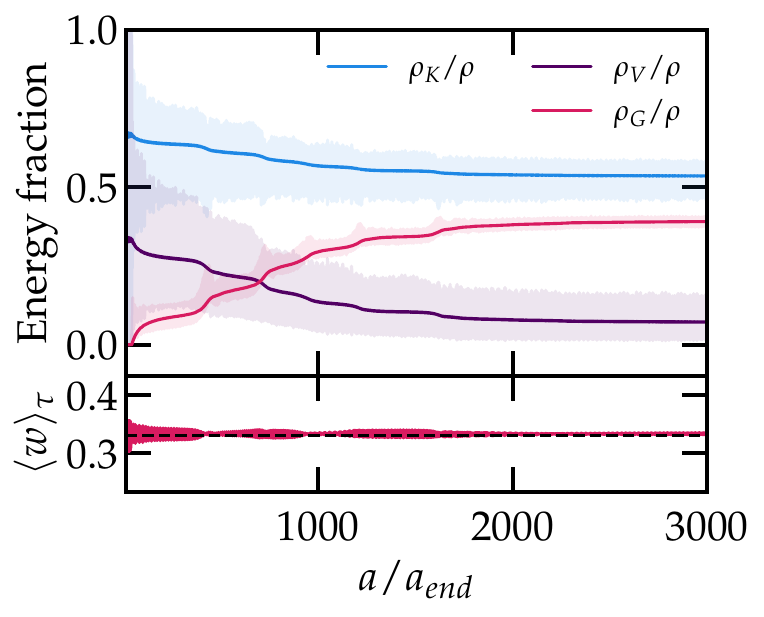} 
\includegraphics[width=0.48\textwidth]{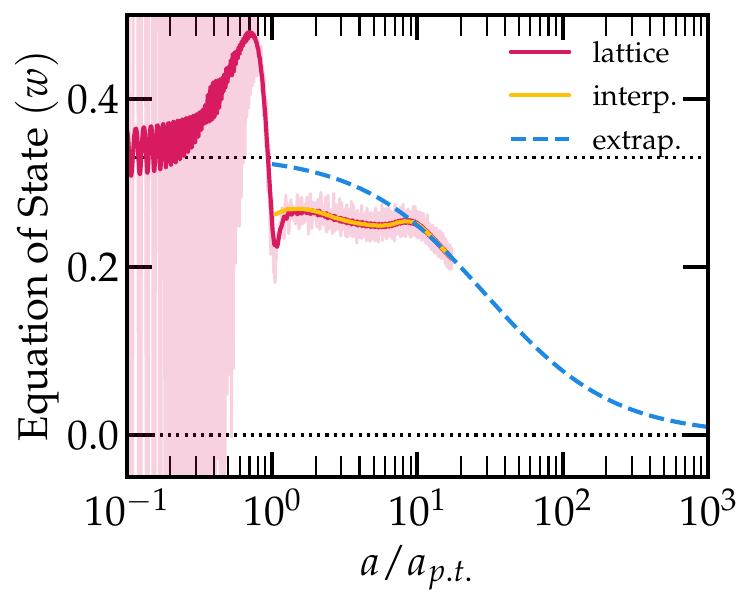}
\caption{\label{fig:energiesprePT} \textit{Upper left}: the kinetic ($\rho_K$), gradient ($\rho_G$) and potential ($\rho_V$) energy fractions (of the total $\rho$) are plotted over time (transparent), along with their time-averages (opaque), which approach constant values by the end of the simulation. (Note that the residual $\rho_V$ is likely subject to finite-size effects.) \textit{Lower left}: the time-averaged equation of state is consistent with radiation ($w \simeq \frac{1}{3}$), even after fragmentation. \textit{Right}: we plot the time evolution of the equation of state, from a simulation with $v_D = 3\times 10^{16}$ GeV ($k_{IR} = 0.08 \sqrt{\lambda_\Phi}\varphi_{\text{end}}$, $N= 128$, $dt = 10^{-3}\sqrt{\lambda_\Phi}\varphi_{\text{end}}$), during and after the non-thermal phase transition, together with a natural interpolation in yellow (leading to an extrapolation in blue) explained in the text. We rescale with a fiducial scale factor in order to use these fits for smaller $v_D$ consistent with (\ref{eq:paramspace}). }
\end{figure}

The backreaction is nicely illustrated if one considers an uncontaminated ``background" energy density 
$\rho_{bg} = \frac{1}{2} \dot\varphi^2 + \frac{\lambda_{\Phi}}{4} (\varphi^2 - v_D^2)^2$,
so that the remainder of the energy density 
will then be associated to fluctuations~\cite{Garcia:2023eol}.
Then, as can be seen in Figure \ref{fig:lattice}, the preheating epoch ends when $\rho_{\text{bg}} \sim \delta\rho$, and the coherent inflaton oscillations are ``fragmented" into an inhomogeneous aggregate of non-thermal decay products with a phase-space distribution function $f(k,t)\simeq n_k(t)$ inferred from the occupation number of $k$ modes of the perturbed fields. By this time, ``re-scattering'' of modes lying in the instability bands has smeared the particle distributions into a highly-occupied phase-space shell $m_T < k < k_{\text{tail}} \sim O(10)\sqrt{\lambda_\Phi}\tilde\varphi_{\text{end}}$, see Figure \ref{fig:lattice}, where modes are excited into a classical regime. In contrast, ultraviolet modes $k > k_\text{tail}$ remain in vacuum. 

Although the resulting non-thermal distribution is peaked in the infrared ($n_k \propto k^{-3/2}$ as can be seen from~\cite{Micha:2002ey}), the particle masses are still red-shifting as $\propto a^{-1}$, and the radiation equation of state is smoothly preserved even after the inflaton condensate becomes subdominant (see Figure \ref{fig:energiesprePT}).
While the relativistic decay products are generally low-energy in comparison to a thermal distribution ($n_k \propto k^{-1}$), suppressing their annihilation rate into DM and SM, the non-thermal fluctuations $\langle \delta \phi^2 \rangle$ are much larger than the thermal $\sim T^2$ fluctuations were the inflaton  instantaneously thermalise~\cite{Kofman:1995fi,Rajantie:2000fd,Kolb:1996jr},
and restore a minimum at the field-space origin for comparatively large values of $v_D$~\cite{Tkachev:1998dc}. 

The system eventually reaches a stationary regime~\cite{Figueroa:2016wxr} where $2\rightarrow 2$ scatterings are dominant, the energy components satisfy certain equipartition relations and the fluctuation energies approach a constant value (see Figure \ref{fig:lattice}). 
Nonetheless, the particle distributions continue to evolve in time, as energy is slowly transported in a ``cascade" from the infrared into the under-occupied ultraviolet modes~\cite{Micha:2002ey,Micha:2004bv}, and this process is expected to continue until the distributions are rendered thermal (a much longer time-scale than what we consider).
This process exhibits turbulence~\cite{Micha:2002ey,Micha:2004bv}, and the particle distributions evolve in a self-similar way after some time, say $\tau_{\text{turb}}$, for indices $p, q$, such that, 
\begin{equation}\label{eq:selfsimilar}
    n_{k}(\tau) = \left(\frac{\tau}{\tau_\text{turb}}\right)^{-q} n_{\tilde{k}}(\tau_\text{turb}),\quad \text{with}\quad \tilde{k} = \left(\frac{\tau}{\tau_\text{turb}}\right)^{-p}k,
\end{equation}
resulting in characteristic power laws for the field variances $\langle a^2\delta\phi_{R,I}^2 \rangle \propto \tau^{-2p}$ which, together with the effective masses, fall slightly faster than due to red-shifting alone. This also implies that the peak fluctuation energy during the energy cascade evolves as $\omega^{\text{peak}}_k (\tau) \sim \left(\tau/\tau_\text{turb} \right)^p \omega^{\text{peak}}_k (\tau_\text{turb})$
and we use this to approximate the time dilation factor for the perturbative decays studied in Section \ref{sec:sec3Bb} (see Appendix~\ref{sec:pertdecay}). During the stationary regime reached by the end of our simulations, we infer $p\simeq0.27$ and $q\simeq 3.5p$ which roughly match with Ref.~\cite{Micha:2002ey}, 
corresponding to $2\rightarrow 2$ scatterings becoming dominant by the end of the simulation, which we assume to be observed until fluctuation masses become important. 

\paragraph{Non-thermal phase transition and $Z'$ domination} 

Eventually, around $a_{p.t.} \sim \tilde{\varphi}_{\text{end}}/v_D$, the inhomogeneous scalar field oscillation amplitude red-shifts into a regime comparable to $v_D$, and the non-trivial vacuum manifold of the effective potential is redeveloped; the minimum having earlier been restored to the field-space origin 
\cite{Tkachev:1995md,Tkachev:1998dc,Kasuya:1997ha,Kasuya:1998td,Kofman:1995fi,Khlebnikov:1998sz}. Once the energy density is sufficiently reduced, the field oscillations are confined to the quadratic valley encircling the unstable maximum at the field-space origin. Following~\cite{Rajantie:2000fd,Khlebnikov:1998sz}, for $g_D^2 \ll \lambda_\Phi$, we do not expect disconnected minima; nevertheless, the correlation length has been reduced to a sub-Hubble regime by inflaton fragmentation and rescattering~\cite{Bassett:1999mt,Rajantie:2000fd,Kaya:2009yr,Lozanov:2019jff}.  As a result, (near-global) topological string loops develop within the lattice box \cite{Tkachev:1998dc,Kasuya:1998td,Dufaux:2010cf,Lozanov:2019jff}, dissolving into both Goldstone (relativistic $Z'$) and massive radial modes as they evolve and intercommute~\cite{Saurabh:2020pqe,Gorghetto:2018myk,Saikawa:2024bta, Baeza-Ballesteros:2023say,Baeza-Ballesteros:2025spb}.\footnote{Even though the string formation is different to the Kibble-Zurek mechanism~\cite{Kibble:1976sj,Zurek:1985qw,Zurek:1993ek}, we may consider the consequences of the expected formation of a string network which follows the usual scaling regime at late times. This would allow for the emission of a \textit{sub-dominant} population of relativistic $Z'$, peaked at $k\sim H$ (but extending up to the string-core scale~\cite{Gorghetto:2018myk}), and that this continues until $H < g_D v_D$~\cite{Long:2019lwl}. Thereafter, the production of $Z'$ experiences kinematic suppression, and we assume that it can be neglected at later times. As reheating ends when $\Gamma_{Z'} \sim H$ and $\Gamma_{Z'} < g_Dv_D$, this is automatically satisfied before the end of reheating in our setup. Under these assumptions, our results in Section \ref{sec:sec3Bb} are unaffected. }
(The emission of the transverse modes are suppressed due to $g_D^2 \ll \lambda_\Phi$~\cite{Long:2019lwl}.) 

As their effective mass now no longer red-shifts with the physical momenta, and rescattering into relativistic radial modes eventually becomes kinematically suppressed, the non-relativistic, $k<m_\rho$, radial modes come to dominate the energy density, resulting in a gradual transition to a matter-like equation of state and a suppression of the energy fraction in Goldstone modes. Our simulations only captured the initial stages of this process (see Figure \ref{fig:energiesprePT}), due to the long time-scale required, which would be shorter if $g_D \gtrsim \lambda_\Phi$. 
A fairly accurate interpolation for the total equation of state (yellow in Figure~\ref{fig:energiesprePT}) is given by $w = \frac{\rho_G}{3(\rho_G + \rho_V)}$, assuming $\rho_K = \rho_G + \rho_V$; leading to a natural extrapolation, using $\rho_V \propto a^{-3}$, $\rho_G \propto a^{-4}$ with their final simulated values, that fits well at late times (dashed blue).\footnote{After a sufficient time, the fluid comprises a decoupled radiation-like component (contributing all of the gradient energy with $(\rho_K)_{\text{rad}} = \rho_G$) and a matter-like component (contributing all of the the potential energy with $(\rho_K)_{\text{mat}} = \rho_V$), so that $\rho_K = (\rho_K)_{\text{rad}} + (\rho_K)_{\text{mat}} = \rho_G + \rho_V$.} We can estimate the amount of expansion before matter-like scaling of the radial modes, by taking the ratio of the energy spectrum peak to the mass, $k_{\text{peak},\rho}/m_{\rho} \sim O(10)$. As can be seen in Figure~\ref{fig:energiesprePT}, this is around where the yellow and blue contours begin to converge.

At some later time $t_{d}$, and for sufficiently small $\lambda_{\Phi H}$, the \textit{dominant} energy fraction in radial modes must be transferred into $Z'$ as Goldstone radiation (since $m_\rho \gg m_T$). The depletion of the non-relativistic radial modes is made possible by cubic interactions generated at tree-level by $\langle \Phi \rangle = v_D$. 
We may obtain an upper-bound on $t_d$ by using the perturbative decay rate for the inflaton, since other non-perturbative effects, such as parametric resonance or Bose enhancement, will only speed up this process if, indeed, they are relevant. Accordingly, we have
\begin{equation}\label{eq:inflifetime}
    \frac{a_{d}}{a_{p.t.}} \lesssim 90\  \left( \frac{2.5\times10^{-10}}{\lambda_\Phi} \right)^{2/3} \left( \frac{v_D}{10^9\ \text{GeV}}\right)^{2/3} 
\end{equation}
and we deduce that the inflaton matter-domination is quite insignificant (compare Figure~\ref{fig:energiesprePT}). After this moment, the dominant energy component is longitudinal $Z'$ with typical momenta $\sim m_\rho/2$.  The end of reheating (and the freeze-in of DM from the relic population of $Z'$) can then be studied perturbatively using coupled Boltzmann equations, as we do below.

\paragraph{Long-wavelength perturbations} A bosonic reheaton with sub-Hubble mass during inflation raises an interesting question about the transferral of inflationary power spectra to radiation and matter components: does the reheating epoch modify the predictions of the inflation model discussed in Section \ref{sec:sec3A}? 
We noted in Section \ref{sec:sec3A}, that the $k\rightarrow 0$ perturbations in the energy density stored in the longitudinal (and transverse) components of the $Z'$ are suppressed. As the effective mass term approaches the transverse mass at long wavelengths, and $g_D^2 \ll \lambda_\Phi$, these modes are not amplified by parametric resonance, or tachyonic instability, at preheating, which we also confirmed numerically by solving the $k\rightarrow 0$ mode equations. Accordingly, before the inflaton decay to $Z'$ becomes efficient, the long-wavelength $Z'$ curvature perturbation is $\zeta_{Z'} \sim 0$. Then, analogously to standard inflaton $\rightarrow$ radiation reheating scenario~\cite{Riotto:2002yw}, the $Z'$ inherits the curvature perturbation as the inflaton energy density is transferred to $Z'$. (In other words, $\zeta = \zeta_\rho$ is a fixed point of the energy transfer from inflatons to unperturbed radiation at $k\rightarrow 0$). The process then repeats as the $Z'$ again decays to $\chi$ and SM radiation. In this way, the proposed reheaton mechanism never generates a large isocurvature perturbation at long wavelengths, and our single-field extrapolation for the CMB spectra is self-consistent.

\subsubsection*{Perturbative reheating era}
\label{sec:sec3Bb}

In order to study the end of reheating and the freeze-in production of DM, we solve the coupled set of integrated Boltzmann equations (\ref{eq:pertevol1}-\ref{eq:pertevol4}) for the inflaton, $Z'$, SM and DM energy densities ($\rho_{\delta\phi}$, $\rho_{Z^\prime}$, $\rho_R$ and $\rho_\chi$) using perturbative approximations for the energy transfer rates, dominated by decays (supplied in Appendix~\ref{sec:pertdecay}). While this approximation may describe the DM and SM bath evolution relatively well in principle, it does not accurately describe the inflaton and $Z'$ bath evolution driven by non-perturbative effects. Hence, in addition to initial conditions (\ref{eq:ics}), we use scaling relations for the effective masses (\ref{eq:mass1}-\ref{eq:mass3}), equations of state $w$ (\ref{eq:eos1}-\ref{eq:eos2}) and time-dilation factors $\gamma$ (\ref{eq:timedil1}-\ref{eq:timedil2}) for the decay source terms, motivated by the results of our lattice simulations discussed in  Section \ref{sec:sec3Ba}. Moreover, as DM is produced by non-thermal reheaton decays with $\Gamma_{Z^\prime \rightarrow \chi\bar{\chi}} \ll H$, we also handle the distribution function with some care (see Appendix \ref{sec:appDM}), so that the transition to matter-like scaling, and hence the DM yield, is more accurately predicted.

As discussed earlier, after preheating, but before the phase transition, the system eventually reaches a stationary state comprising inflaton and longitudinal $Z'$ fluctuations, and a subdominant inflaton background component (see Figure~\ref{fig:lattice} bottom-right) drained after preheating (which we hence subsume into $\rho_{\delta\phi}$ below). For, $v_D < 10^{12}$~GeV, we obtain final values from the lattice in this stationary regime as the initial conditions for the subsequent bath evolution (\ref{eq:ics}), using subscript ``s'' to denote this time. (The $U(1)_D$ phase transition later takes place before reheating has completed, and we ensured that relevant effects of the non-linear field evolution, inferred in the previous section for larger $v_D$, were properly incorporated.) The dynamics of the subsequent energy transfer are then well-approximated by solving:
\begin{eqnarray}
\label{eq:pertevol1}
\frac{\mathrm{d}\rho_{\delta\phi}}{\mathrm{d}N} + 3\,(1+w_{\delta\phi})\rho_{\delta\phi} &=& -\frac{\left(\gamma_\phi \Gamma_{\delta\phi \rightarrow hh} +  \gamma_\phi \Gamma_{\delta\phi \rightarrow Z^\prime Z^\prime} \right)}{H}\,(1+w_{\delta\phi})\rho_{\delta\phi} 
\\
\label{eq:pertevol2}
\frac{\mathrm{d}\rho_{R}}{\mathrm{d}N} + 4\,\rho_{R} 
&=& \frac{\gamma_\phi \Gamma_{\delta\phi \rightarrow hh}}{H}\,\,(1+w_{\delta\phi})\rho_{\delta\phi} 
+\frac{\gamma_{Z^\prime}\Gamma_{Z^{\prime} \rightarrow {\rm SM}{\rm SM}}}{H}\,(1+w_{Z^\prime})\,\rho_{Z^\prime},
\\ 
\label{eq:pertevol3}
\frac{\mathrm{d}\rho_{Z^\prime}}{\mathrm{d}N} + 3\,(1+w_{Z^\prime})\,\rho_{Z^\prime} &=& \frac{\gamma_\phi \Gamma_{\delta\phi \rightarrow Z^\prime Z^\prime}}{H}\,(1+w_{\delta\phi})\rho_{\delta\phi} \nonumber
\\
&& - \left(\frac{\gamma_{Z^\prime}\Gamma_{Z^{\prime} \rightarrow {\rm SM}{\rm SM}}}{H} + \frac{\gamma_{Z^\prime}\Gamma_{Z^\prime \rightarrow \chi\bar{\chi}}}{H}\right)\,(1+w_{Z^\prime})\,\rho_{Z^\prime},
\\
\label{eq:pertevol4}
\frac{\mathrm{d}\rho_{\chi}}{\mathrm{d}N} + 3\,(1+w_{\chi})\rho_{\chi} &=& \frac{\gamma_{Z^\prime}\Gamma_{Z^\prime \rightarrow \chi\bar{\chi}}}{H}\,(1+w_{Z^\prime})\,\rho_{Z^\prime},\\
\label{eq:pertevol5}
H^2 &=& \frac{\rho_{\delta\phi} + \rho_{Z^\prime} + \rho_R + \rho_\chi}{3M_P^2};
\end{eqnarray}
with all the relevant expressions for $w$, $\gamma$ and $\Gamma$ given in Appendices~\ref{sec:pertdecay} and ~\ref{sec:appDM}.\footnote{We assume that Bose enhancement and Pauli blocking effects are not relevant.} The non-thermal relic density is then inferred from $Y_\infty$, well after the $\chi$ has become non-relativistic. 

\begin{figure}[t!]
\centering
\includegraphics[width=0.85\textwidth]{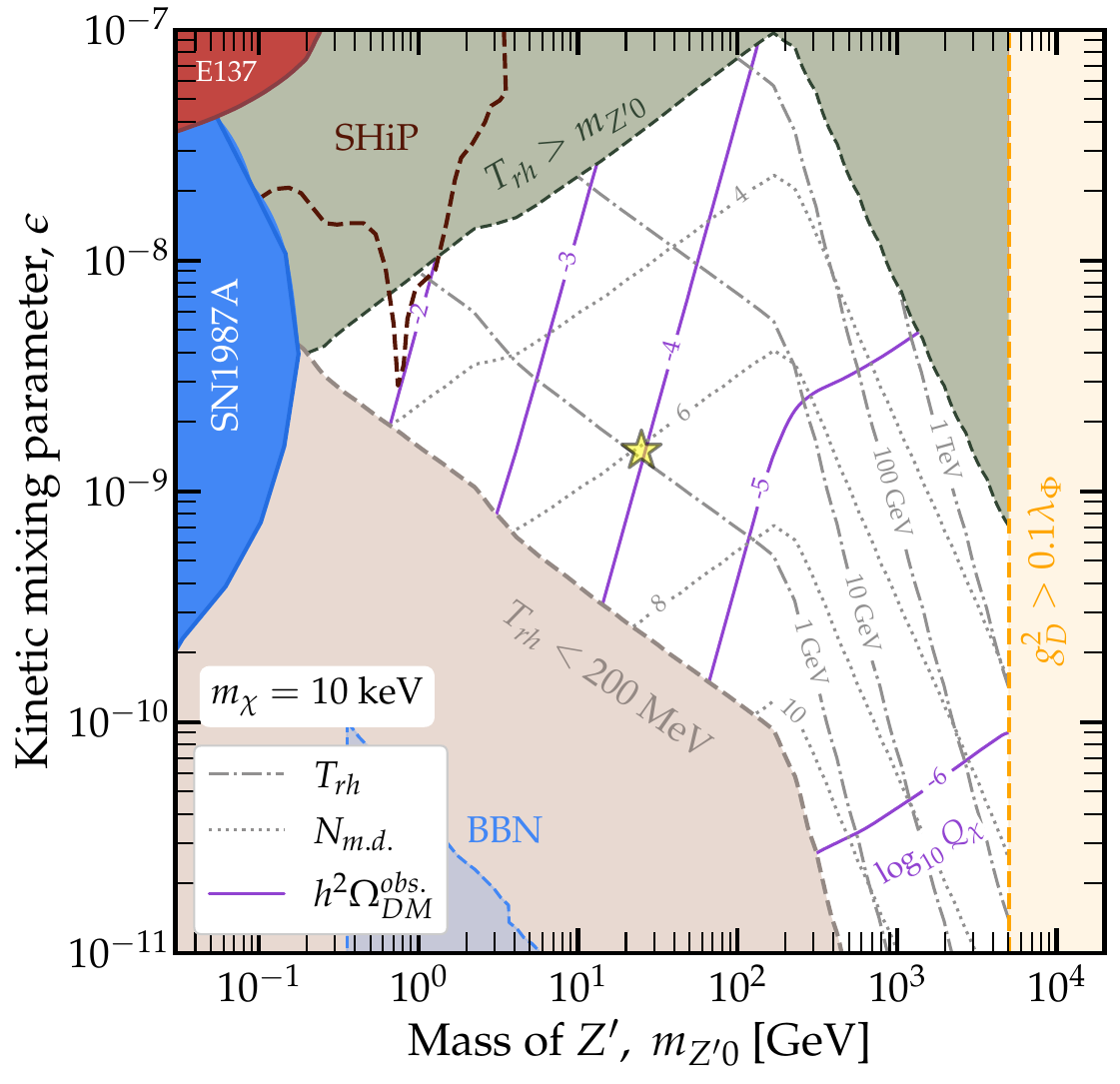}
\caption{\label{fig:relicdensity} Here we present the region of parameter space spanned by $\epsilon-m_{Z^\prime,0}$ where our proposed cosmology encompassing a $Z^\prime$ reheaton can be achieved (in white). In the green shaded region at the top, the $Z^\prime$ thermalise with the SM plasma, so that freeze-in is instead dominantly thermal. The brown shaded region is properly described by chiral perturbation theory (beyond the scope of this work). Several existing constraints from SN1987A~\cite{Chang:2016ntp}, E137~\cite{Marsicano:2018krp} and BBN~\cite{Berger:2016vxi} are shown by the blue, red and light-blue shaded regions, respectively. The projected sensitivity reach of SHiP~\cite{SHiP:2020vbd} is also shown by the dark red dashed line. The light yellow shaded vertical region on the right indicates the parameter region where the near-global limit of $U(1)_D$ is invalid. Within the allowed region the purple contours represent the contours along which DM freeze-in relic density is satisfied for different values of the DM $U(1)_D$ charge $Q_\chi$, the gray dash-dotted lines denote the contours of different reheating temperatures, while the gray dotted lines represent the contours of different values of $N_{m.d.}$. This plot is obtained for representative values of $v_D = 10^9\,{\rm GeV}$ and $m_\chi = 10\,{\rm keV}$, while the yellow star marks a suitably chosen benchmark point used in our discussions. See the text for details.}
\end{figure}

\begin{figure}[t!]
\centering
\includegraphics[width=0.6\textwidth]{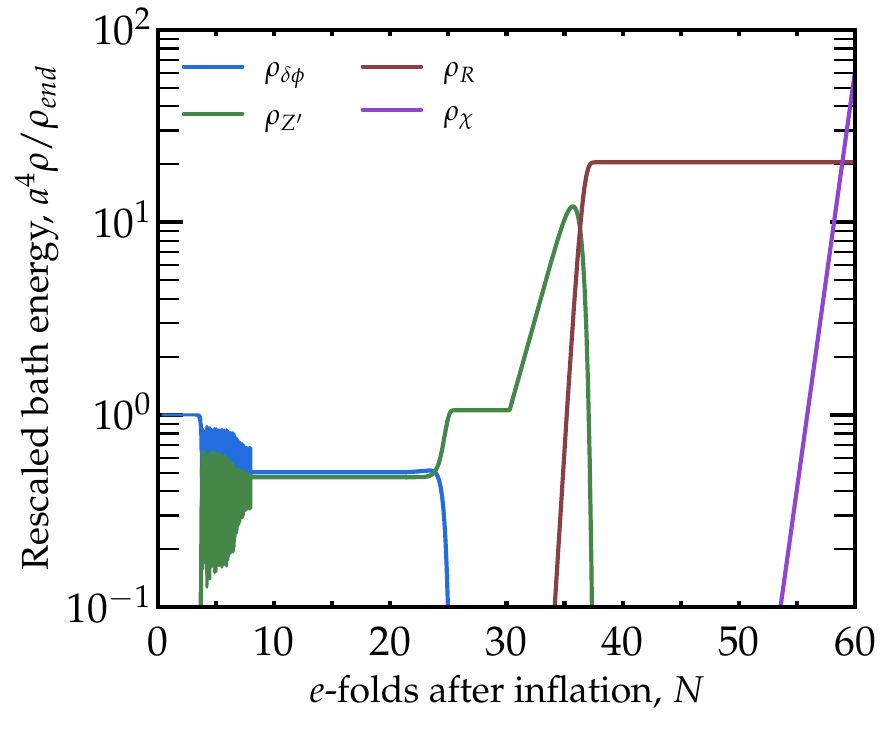}
\caption{
\label{fig:reheatingplot} We plot the example bath evolution, rescaled so that radiation is constant, for the starred benchmark point in Figure \ref{fig:relicdensity}.}
\end{figure}

In Figure~\ref{fig:reheatingplot}, we plot an example solution to (\ref{eq:pertevol1}-\ref{eq:pertevol4}), along with the initial lattice output, for the benchmark point shown by yellow star in Figure~\ref{fig:relicdensity} ($m_{Z^\prime,0} = 25$ GeV, $\epsilon = 10^{-9}$, $Q_\chi \sim 10^{-4}$).
One sees that the energy budget becomes dominated by $Z^\prime$s relatively quickly after the non thermal phase transition (compare \ref{eq:inflifetime}). Relativistic at production, the $Z^\prime$s from inflaton decay then red-shift into a non-relativistic regime after $\sim 30$ e-folds. The universe then enters a matter-dominated era, lasting for $\sim 6$ e-folds, before the initially $T_{rh}\sim 1$ GeV radiation bath is generated by $Z^\prime \rightarrow {\rm SM}\,{\rm SM}$ decays. During the $Z'$ domination, a sub-dominant bath of $\chi$ is produced, which reach a matter-like equation of state in time to reproduce the relic density $\Omega_\chi h^2 = 0.12$. 

In Figure~\ref{fig:relicdensity}, we delineate 
the $\epsilon-m_{Z^\prime,0}$ parameter space, for 
$v_D = 10^{9}\,{\rm GeV}$, where a non-thermal $Z^\prime$-dominated 
cosmological era exists prior to the thermal radiation dominated era with 
initial reheating temperature $T_{\text{rh}}$ defined by $\rho_{Z^\prime}=\rho_R$. 
The vertical yellow band on the right describes the maximal assumed value of 
$m_{Z^\prime,0}$ above which the near-global limit is no longer satisfied, and the perturbative inflaton decay to $Z^\prime$, which leads to reheaton domination in our setup, becomes kinematically prohibited. (A smaller $v_D$ simply shifts this line to the left.)
The upper green region represents the parameter space where the $Z^\prime$s thermalise with the resulting SM bath at the end of reheating either through annihilations or inverse decays (corresponding to $T_{rh} > m_{Z^\prime,0}$). Consequently, the predicted DM production rate in this region is presumably well-described by earlier thermal freeze-in studies, and is not a region of interest. (This requirement is more restrictive than the demand for primordial $Z^\prime$ domination.)
The light brown region in the lower left side of the plot merely depicts the region where $T_{rh} < \Lambda_{\rm QCD}$ and a proper treatment 
of $Z^\prime$ decays to SM needs careful invocation of Chiral Perturbation 
Theory. (Of course, within this region, the parameter space is ultimately enclosed from below by the requirement of sufficiently high reheating temperatures for BBN.)

The remaining substantial white space shows the parameter region where one realises non-thermal freeze-in DM production from $Z^\prime$ reheatons. (We have also taken into account the SM 
contributions via ${\rm SM}\,{\rm SM} \rightarrow \chi\bar{\chi}$ processes 
which are much smaller than the Hubble expansion rate for all of our 
parameter choices.) 
Within this region, we draw contours which reproduce the expected relic density for different DM charges $Q_\chi$ in purple, dotted lines denote the e-folds of matter-domination (which are linked to the inflationary predictions in Figure \ref{fig:rnsplot}), while dot-dashed lines provide the corresponding reheating temperatures, which are lower than those compatible with standard thermal freeze-in at the same ($m_{Z^\prime,0}, \epsilon$).

We additionally show, for reference, the constraints coming from the observations of SN1987A energy 
losses~\cite{Chang:2016ntp} and from the electron beam-dump experiment 
E137~\cite{Marsicano:2018krp}. The excluded parameters are given 
by the blue-shaded and the red-shaded regions, respectively. 
Moreover, the sensitivity of the upcoming 
SHiP~\cite{SHiP:2020vbd} observations 
for displaced decays of $Z^\prime$s are shown by a dashed red curve, probing some of the available parameter space for $m_{Z^{\prime},0} \sim 0.6-1$ GeV and $\epsilon \sim$ few $\times 10^{-9}$. The lower light blue patch, corresponding to lower reheating temperatures than we explicitly study, represents a parameter 
region ruled-out due to the over production of 
$^4 {\rm He}$ and $^2{\rm H}$~\cite{Berger:2016vxi}.

\section{Gravitational waves}
\label{sec:sec4}

\begin{figure}[htb!]
\centering 
\includegraphics[width=0.65\textwidth]{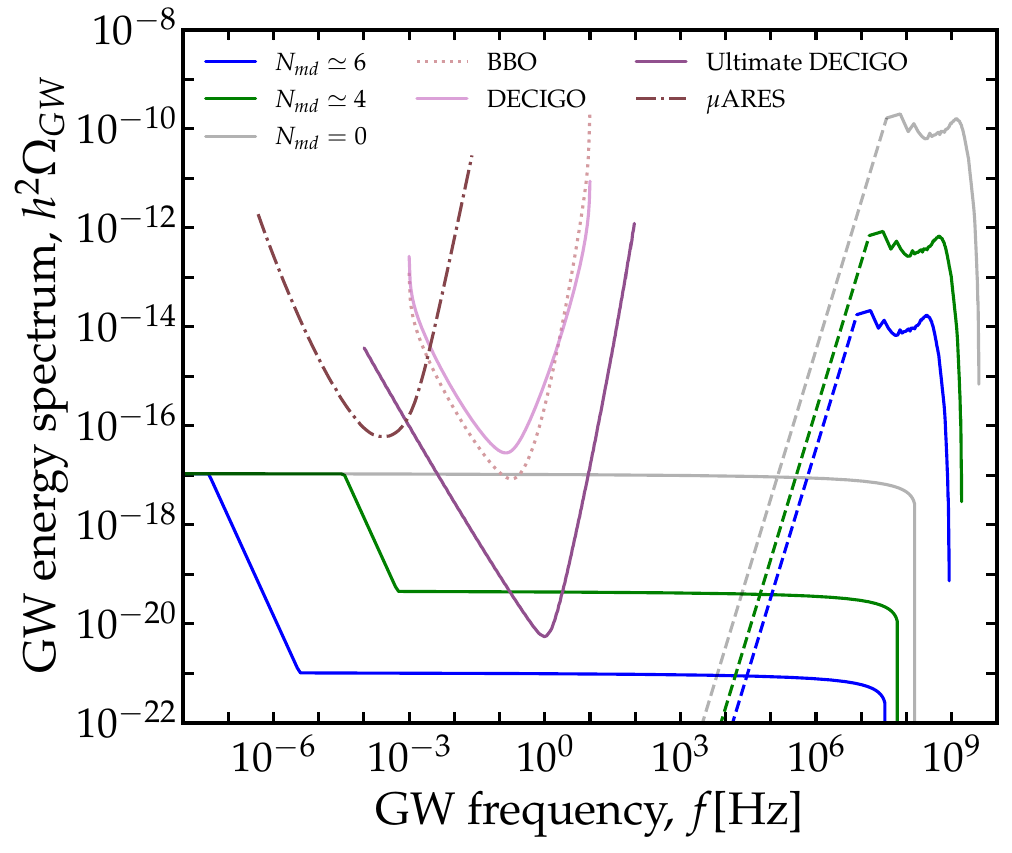}
\caption{\label{fig:gwomega} We include representative contours (explained in the text) for the SGWB contributed by the inflation and preheating epochs in the model for the starred benchmark point in Figure \ref{fig:relicdensity} (blue) which, together with a $T_{rh} \sim$ TeV, $N_{m.d.} \sim 4$ parameter point (green), are compared to the no $Z^\prime$ reheaton scenario in grey; along with the projected sensitivity curves for BBO~\cite{Crowder:2005nr}, DECIGO~\cite{Seto:2001qf} (obtained from Refs.~\cite{Schmitz:2020syl,schmitz_new_2020}), as well as $\mu$~\cite{Sesana_2021} and ultimate DECIGO~\cite{Kuroyanagi_2015} sensitivities. A smaller choice of $\xi_\Phi$ than our benchmark value will result in a larger amplitude inflationary spectrum than what is plotted. } 
\end{figure}

In this section, we provide some estimates for the two most significant contributions to the high-frequency stochastic gravitational wave background (SGWB) corresponding to the cosmic history above; our main result is summarised in Figure \ref{fig:gwomega}. Direct detection at these frequencies remains challenging, but may be achieved in principle with the prospective space-borne laser interferometers: DECIGO~\cite{Seto:2001qf} (particularly at ``ultimate'' sensitivity~\cite{Kuroyanagi_2015}), BBO~\cite{Crowder:2005nr} and potentially $\mu$Ares~\cite{Sesana_2021} for the smaller regime of $\xi_\Phi$. As we explain below, we can expect a kink in the spectrum at frequencies  $\lesssim 10^{-4}$ Hz associated to the reheating temperature $< \mathcal{O}({\rm TeV})$, and a suppression at $\sim$Hz frequencies proportional to $N_{m.d.}$; identifying a single parameter point in Figure \ref{fig:relicdensity}, if measurable. While the former is likely below the threshold of proposed $\mu$-mHz experiments such as $\mu$Ares, the latter is detectable in principle for $N_{m.d.} >10$.

Initially super-horizon tensor perturbation modes, generated with an approximately flat power spectrum $\Delta^2_{t} (k, \tau_i)$ during the inflationary epoch, eventually comprise an irreducible contribution to the cosmological SGWB~\cite{Grishchuk:1974ny,Starobinsky:1979ty}; with an amplitude attenuated across different frequency ranges by the expansion history. For modes entering the thermal radiation era, $k = 2\pi a_0f \in (a_{\text{eq}}H_{\text{eq}}, a_{\text{reh}}H_{\text{reh}})$, with a horizon crossing temperature $T_{\text{hc}}$ it may be shown (see Refs.~\cite{Saikawa:2018rcs,Ringwald:2020vei}) that the fraction of the critical density per logarithmic frequency interval is
\begin{equation}
\begin{split}
    h^2 \Omega^{\text{inf,RD}}_{\text{GW}} (f) 
    &= \frac{1}{\rho_c}\frac{d\ln \rho_{\rm GW}}{d\ln k} \simeq 1 \times 10^{-16}\times \left[ g_{*s}(T_{\text{hc}})\right]^{-4/3} g_{*\rho}(T_{\text{hc}}) \left [ \frac{H_*(f)}{3\times10^{13}\ \text{GeV}}\right]^{2}, \\
    \text{where}\quad f &\simeq 1\times10^{-5}\ \text{Hz}\ \times \left[ g_{*s}(T_{\text{hc}})\right]^{-2/3} [g_{*\rho}(T_{\text{hc}})]^{1/2}\left[\frac{T_{\text{hc}}}{1\ \text{TeV}} \right];
\end{split}
\end{equation}
with the last expression giving $T_{\text{hc}}$ implicitly in terms of the present frequency, $f$, while `$*$' denotes the point of horizon exit. On the other hand, for modes entering the horizon during the reheating epoch with equation of state $w$~\cite{Zhao:2011bg,Liu:2015psa}), the corresponding quantity is:
\begin{equation}\label{eq:matterdomgw}
    h^2 \Omega^{\text{inf, pre-RD}}_{\text{GW}} (f) \simeq \left(\frac{k_{\text{rh}}}{k}\right)^{\frac{2(1-3w)}{1+3w}} \times 10^{-16}\times \left[ g_{*s}(T_{\text{rh}})\right]^{-4/3} g_{*\rho}(T_{\text{rh}}) \left [ \frac{H_*(f)}{3\times10^{13}\ \text{GeV}}\right]^{2}.
\end{equation}
We may then set $w=0$ for early matter-domination, which is subjected to $\propto k^{-2}$ damping at high frequencies. These expressions are used to draw the broader and flatter contours in Figure \ref{fig:gwomega}. It may be seen that the matter-like reheaton leaves a distinctive feature in the spectrum.

Additionally, as we saw in Section \ref{sec:sec3Ba}, large sub-horizon field gradients are developed in both the inflaton and the Goldstone fields during preheating. This results in an extensive production of GWs~\cite{Khlebnikov:1997di,Easther:2006gt,Easther:2006vd,Dufaux:2007pt,Garcia-Bellido:2007nns,Garcia-Bellido:2007fiu,Easther:2007vj,Dufaux:2008dn,Figueroa:2017vfa,Caprini:2018mtu} with additional ultraviolet growth during turbulence, only saturating once the system reaches a stationary state. This picture was confirmed in our lattice simulations with $\mathcal{C}$osmo$\mathcal{L}$attice~\cite{Figueroa:2020rrl,Figueroa:2021yhd}, from which we obtain a prediction for the corresponding SGWB fraction
\begin{equation}
    h^2 \Omega^{\text{preh}}_{\text{GW}} (f) \simeq 7.7 \times 10^{-5}\times \left[ g_{*s}(T_{\text{rh}})\right]^{-4/3} g_{*\rho}(T_{\text{rh}}) e^{-N_{\text{m.d.}}}\left\langle\frac{1}{\rho(\tau_\text{turb})} \frac{\mathrm d \rho_{\text{GW}}}{\mathrm d \log k}(\tau_\text{turb}) \right\rangle\Bigg|_{k=\frac{f}{2\pi a_o}}
\end{equation}
where the early matter domination, occurring after production, dampens the spectrum. The spatially averaged quantity is obtained from simulations, and we extrapolate a $\propto k^3$ causal suppression on super-horizon scales at the production time~\cite{Caprini:2018mtu}. The present day contribution from a typical simulated spectral amplitude is plotted to the upper right in Figure \ref{fig:gwomega}.

The SGWB owing to a relic cosmic string network depends on the string tension set by $v_D$, viz. $G\mu \simeq 2\pi G v_D^2\log\left(\frac{m_r}{\Lambda}\right)$ where $\Lambda = \max\{g_D v_D, H \}$~\cite{Kitajima:2022lre}, which is maximised at the end of reheating for
\begin{equation}\label{eq:maximumtension}
    G\mu < 1.1\times10^{-10}\left(\frac{v_D}{10^{13}\ \text{GeV}}\right)^2\left\{1 + \frac{1}{26}\log\left[ \left( \frac{\lambda_\Phi}{2.5\times10^{-10}}\right)^{\frac{1}{2}}\left(\frac{v_D}{10^{13}\ \text{GeV}} \right) \left( \frac{m_{Z^\prime,0}}{2m_e}\right)\right]\right\}
\end{equation}
where $m_e$ is the electron mass. The string tension is constrained by CMB experiments as $G\mu < 2\times 10^{-7}$~\cite{Planck:2013mgr,Lizarraga:2016onn}. A more stringent bound from pulsar timing arrays (PTA) is motivated by simulations using a Nambu-Goto approximation, viz. $G\mu \lesssim 10^{-10}$~\cite{EPTA:2023xxk,NANOGrav:2023hvm,Figueroa:2023zhu}. However, recent dedicated simulations of the Abelian-Higgs model suggest that this bound may be an overestimate, with particle emission suppressing the GW amplitude across all frequencies by many orders of magnitude~\cite{Baeza-Ballesteros:2025spb}.\footnote{Hence, while the SGWB amplitude achieved when saturating (\ref{eq:maximumtension}) in the model may be \textit{in principle} within reach of future experiments (see also Ref.~\cite{Kitajima:2022lre}), given the ongoing uncertainties around the amplitude of the GW signal from the string network, we do not include corresponding contours in Figure~\ref{fig:gwomega}.}
Nonetheless, we note that our regime of $v_D$ (\ref{eq:paramspace}) remains consistent with the most conservative current bounds, as seen from (\ref{eq:maximumtension}).

\section{Summary and Conclusion}
\label{sec:sec5}

In the absence of any conclusive evidence supporting the existence of GeV-TeV scale thermal DM candidates, 
DM particles produced via non-thermal mechanisms are assuming increasing importance. 

To this end, we studied 
the freeze-in production of light fermionic DM particles ($\chi$) within the framework of a 
well-motivated kinetic-mixing portal fermionic DM model. Going beyond the standard approach, where the 
Abelian $Z^\prime$ is assumed to be thermal, we considered cosmic inflation to take place via the slow-roll 
of a dark sector scalar field $\Phi$ such that post-inflationary oscillations of $\Phi$ give rise 
to a bath predominantly consisting of non-thermal $Z^\prime$ in the near-global limit of the underlying 
$U(1)_D$ symmetry, i.e., for $g^2_D \ll \lambda_\Phi$. We additionally required that the SM Higgs 
coupling with the inflaton $\Phi$ was sufficiently weak to ensure that the SM bath is not produced right after the 
inflationary reheating era ends. 

The model features gauge kinetic mixing that drives the decay of the non-thermal
$Z^\prime$ into both the SM particles as well as DM $\chi$ via perturbative decays, leading 
to the SM radiation bath  prior to BBN and requisite DM abundance prior to the thermal era. This implies that the Abelian vector boson 
$Z^\prime$ behaves as a reheaton, a novel possibility. For some typical choice of model parameters, 
we found that the onset of the SM radiation bath occurs at reheating temperatures 
$T_{\rm rh} \in 200\,{\rm MeV} - 10\,{\rm TeV}$ in our scenario and the reheaton 
$Z^\prime$ behaves as a matter-like component for a substantial period thereby slowing 
down the expansion of the Universe. We then showed that this long matter-dominated era imprints a characteristic feature in the 
inflation and preheating induced stochastic gravitational wave background (SGWB) for 
$f > 10^{-8}\ {\rm Hz}$ which can in principle be probed by future space-based 
interferometer observations. 

The importance of this study not only lies in identifying that an alternative yet viable 
cosmological evolution driven by a $Z^\prime$ reheaton can exist within the 
well-studied kinetic-mixing portal fermionic DM model, but also in pointing out the 
novel spectral shape of the SGWB this scenario predicts.

\section*{Acknowledgements}\label{sec:Acknowledgements}
AHS would like to thank Carlos Tamarit for helpful discussions. This work was supported in part by the Australian
Research Council through the ARC Centre of Excellence
for Dark Matter Particle Physics, CE200100008. AHS is supported by the Australian Government Research Training Program Scholarship initiative.

\appendix
\section{Details of perturbative reheating analysis}
\label{sec:pertdecay}

In this Appendix, we provide the 
decay rates, effective masses, time dilation factors, equations of state and initial conditions used in (\ref{eq:pertevol1}-\ref{eq:pertevol4}), motivated by the initial study of the non-perturbative aspects of the reheating in Section \ref{sec:sec3Ba}. 

\paragraph{Inflaton decay rates} The energy dissipation from the inflaton quanta ($\delta\phi$) is dominated by the decay rates:
\begin{eqnarray}
\label{eq:decay1}
\Gamma_{\delta\phi \rightarrow hh} & \simeq & \frac{\lambda^2_{\Phi\,H}v^2_D}{512\pi m_{\phi}} \sqrt{1-\frac{4m^2_{h}}{m^2_{\phi}}},
\\
\label{eq:decay2}
\Gamma_{\delta\phi \rightarrow Z^\prime Z^\prime} & \simeq & \frac{g^4_D v^2_D}{192\pi} \frac{m^3_{\phi}}{m^4_{Z^\prime}} \left[ 1 - 4\frac{m^2_{Z^\prime}}{m^2_{\phi}} + 12\frac{m^4_{Z^\prime}}{m^4_{\phi}} \right] \sqrt{1 - \frac{4m^2_{Z^\prime}}{m^2_{\phi}}},
\end{eqnarray}
where $m^2_{\phi}$, $m^2_{Z^\prime}$ and $m^2_{h}$ are given by (\ref{eq:mass1}-\ref{eq:mass3}).

\vspace{2mm}

\paragraph{$Z'$ decay rates} There exists several possible decay modes of $Z^\prime$ into SM particles which must be added to give the $Z^\prime$ 
total decay rate into SM particles 
$\Gamma_{Z^\prime \rightarrow {\rm SM}{\rm SM}}$. These decay rates are 
tabulated as follows:
\begin{eqnarray}
\Gamma_{Z^\prime \rightarrow W^+W^-}  &=& \frac{\epsilon^2 e^2 c^2_w}{16\pi} m_{Z^\prime} \left[7 - \frac{5m^2_{Z^\prime}}{m^2_W} - \frac{12m^2_W}{m^2_{Z^\prime}} + \frac{m^4_{Z^\prime}}{m^4_W}\right] \sqrt{1-\frac{4m^2_W}{m^2_{Z^\prime}}},\\
\Gamma_{Z^\prime \rightarrow f\bar{f}} &=&  \frac{\epsilon^2 e^2 c^2_w Q^2_f N_c}{4\pi}m_{Z^\prime} \left(1+ \frac{2m^2_f}{m^2_{Z^\prime}}\right) \sqrt{1-\frac{4m^2_f}{m^2_{Z^\prime}}},
\end{eqnarray}
where $Q_f$ and $N_c$ denote the charge and color of the daughter 
SM fermion. In addition, the $Z^\prime$ can also decays into DM $\chi$ 
with a decay rate:
\begin{equation}
\Gamma_{Z^\prime \rightarrow \chi\bar{\chi}} =  \frac{g^2_D}{4\pi} Q^2_\chi m_{Z^\prime} \left(1+ \frac{2m^2_\chi}{m^2_{Z^\prime}}\right) \sqrt{1-\frac{4m^2_\chi}{m^2_{Z^\prime}}}.
\end{equation}

\paragraph{Effective masses}

In addition to the usual SM masses, and the mass parameter $m_\chi$, the decay rates above also depend on the mass-squared of the $Z^\prime$, the inflaton and the Higgs, which are time-dependent until the non-thermal phase transition. All three receive large contributions induced by the coincident 2-point function for the radial field $\langle \rho^2\rangle$ which receives substantial non-thermal corrections which must be included in addition to the background value. Note that, as explained above, the fluctuation contribution redshifts as $\propto \tau^{-2(1+p)}$ during this time. Hence, in solving the bath equations from the remaining evolution after some fiducial time during the stationary non-linear evolution, $\tau_s$, we use
\begin{eqnarray}
    \langle \rho^2 \rangle &\simeq& \max \left \{ \langle \rho^2 \rangle|_{\tau = \tau_s} \left(\frac{a_s}{a} \right)^{2(1+p)} , v_D^2 \right\} \\
    \label{eq:mass1}
    m^2_{\phi} &\simeq& \frac{\lambda_\Phi}{2} \left( 3\langle\rho^2\rangle - v_D^2\right), 
    \\
    \label{eq:mass2}
    m^2_{Z^\prime} &\simeq& g_D^2 \langle\rho^2\rangle + \lambda_\Phi\left(  \langle\rho^2\rangle - v_D^2\right)
    \\
    \label{eq:mass3}
    m^2_h &\simeq& \lambda_{H} v_H^2 +\frac{\lambda_{\Phi H}}{2}(\langle \rho^2 \rangle - v_D^2)
\end{eqnarray}
Note that the second contribution in $m^2_{Z^\prime}$, for which we have used (\ref{eq:effmass}) with large $k$, vanishes after the phase transition (we denote by ``p.t.'' the value at the first moment where $\langle \rho^2 \rangle = v_D^2$). 


\paragraph{Time dilation factors}

The collision terms for the decays in the integrated Boltzmann equations generally lead to a time dilation factor given by
\begin{equation}
    \gamma_I = \left\langle \frac{m_I}{E_I} \right\rangle
\end{equation}
where $\langle\dots \rangle$ here denotes an average over the particle distribution. Before the inflaton decay to $Z^\prime$ become efficient, we make the approximation $\langle E^{-1}\rangle \sim a_s/[ak_{\text{peak}}(\tau)]$ where $k_{\text{peak}}(\tau) = k_\text{peak}(\tau_s)(\frac{a}{a_s})^p$ is the peak of the $\rho_k$ distribution for each species (i.e. the typical momentum). Note that before the phase transition this quantity is  small and approximately constant (as the effective mass is redshifting similarly), but grows towards unity with $a^{p+1}$ afterwards. After the inflaton decays to $Z^\prime$, we take the latter to be monoenergetic with $\langle E^{-1}\rangle \sim m_{\phi}^{-1}$, by assuming these are instantaneous at time ``d'' when $\Gamma_{\delta\phi \rightarrow Z^\prime Z^\prime} = H$.
In summary, we assume the following expressions for $\gamma_\phi$ and $\gamma_{Z^\prime}$
\begin{eqnarray}
    \label{eq:timedil1}
    \gamma_{ \phi} &=& \min \left\{1, \frac{m_{\phi}}{k_{\text{peak},\phi}(\tau_{\text{s}})} \left(\frac{a_s}{a} \right)^{p+1} \right\}  
    \\
     \label{eq:timedil2}
    \gamma_{Z^\prime} &=& 
    \begin{cases}
        & \min \left\{1, \frac{m_{Z^\prime}}{k_{\text{peak},\phi}(\tau_{\text{s}})} \left(\frac{a_s}{a} \right)^{p+1}  \right\} \ ,\qquad a \leq a_{d}
        \\
        & \qquad \min \left\{1, \frac{g_D}{\sqrt{\lambda_{\Phi}}} \frac{a}{a_{d}} \right\} \qquad  \quad ,\qquad a > a_{d}.
    \end{cases}
\end{eqnarray}

\paragraph{Equations of state}

We make the following approximation for the inflaton equation of state 
\begin{equation}
    \label{eq:eos1}
    w_{\delta \phi} = 
    \begin{cases}
        &\qquad \quad \frac{1}{3}\qquad \quad \  \ , \quad a\leq a_{p.t.} \\
        & \frac{(\rho_G)_f}{3[(\rho_G)_f + \frac{a}{a_f}(\rho_V)_f]}\ \ ,\quad a>a_{p.t}
    \end{cases}
\end{equation}
where the latter quantity is plotted in Figure \ref{fig:energiesprePT}, with subscript $f$ denoting the final value of the corresponding simulation. The equation of state for the $Z^\prime$ simply follows from the above discussion of the time dilation factors, for which we use a Heaviside approximation
\begin{equation}
    \label{eq:eos2}
    w_{Z^\prime} = \frac{1}{3}\theta(1 -\gamma_{Z^\prime})
\end{equation}
We use a numerical solution for the dark matter equation of state, see Figure \ref{fig:dmeos}, with details given in \ref{sec:appDM}.

\paragraph{Initial conditions}

In producing Figure~\ref{fig:relicdensity}, we used a lattice simulation with our benchmark values for the first $N_s \simeq 8$ efolds, 
ending (at $\tau_s$) after the stationary regime was apparently reached. We then infer initial conditions for the subsequent bath evolution from the kinetic energy contributions of the $\Phi$ components, and the total SM Higgs energy as follows
\begin{equation}\label{eq:ics}
    \rho_{\delta\phi}(N_s) \simeq \langle \dot\phi_R^2(\tau_s)\rangle,\quad \rho_{Z^\prime}(N_s) \simeq \langle \dot{\delta\phi^2_I}(\tau_s)\rangle,\quad \rho_{R}(N_s) \simeq \rho_{h}(N_S)
\end{equation}
where $\langle \dots \rangle$ here denotes a lattice-averaged quantity. Note that $\rho_{\delta\phi}$ includes the zero mode contribution. We set $\rho_\chi(\tau_s) \simeq 0$ as $Z^\prime$ decays are completely negligible at early times. 

\section{Scattering rates for $Z'$ thermalisation and DM freeze-in}
\label{sec:thermalization}

The necessary scattering rates that must be smaller than the Hubble rate in 
order to achieve a non-thermal Universe prior to SM reheating are as follows:

\begin{enumerate}[\leftmargin=-1cm]
\item \underline{${\rm SM}{\rm SM} \rightarrow Z^\prime Z^\prime$:} These processes 
are $t,u$-channel SM-mediated processes and can thermalise $Z^\prime$s with the 
SM plasma unless the corresponding rates are smaller than $H$ during the 
entire $Z^\prime$-dominated era. The relevant cross-sections are:
\begin{eqnarray}
\langle \sigma_{f\bar{f} \rightarrow Z^\prime Z^\prime}v \rangle &\simeq & 
\begin{cases}
4\pi\epsilon^4 \alpha^2_{\rm EM}\cos^4\theta_W Q^4_f \frac{\left(m^2_f-m^2_{Z^\prime,0}\right)}{(2m^2_f-m^2_{Z^\prime,0})^2}\sqrt{1-\frac{m^2_{Z^\prime,0}}{m^2_f}}\,\,\,{\rm for}\,\,m_f> T,\\
4\pi\epsilon^4 \alpha^2_{\rm EM}\cos^4\theta_W Q^4_f \frac{75}{s}\,\,\,{\rm for}\,\,m_f < T\,\,{\rm and}\,\,s\simeq 9T^2,
\end{cases}
\end{eqnarray}
and
\begin{eqnarray}
&&\langle \sigma_{W^+W^- \rightarrow Z^\prime Z^\prime}v \rangle \simeq  \nonumber\\\!\!\!
&&\begin{cases}
\frac{\pi\epsilon^4 \alpha^2_{\rm EM} \cos^4\theta_W}{9m^2_{W}}\frac{\left(m^2_W-m^2_{Z^\prime,0}\right)^2}{(2m^2_W-m^2_{Z^\prime,0})^2}\sqrt{1-\frac{m^2_{Z^\prime,0}}{m^2_W}}\left( \frac{33m^4_W+58m^2_W m^2_{Z^\prime,0}+3m^4_{Z^\prime,0}}{m^4_{Z^\prime,0}}\right)\,\,{\rm for}\,\,m_W> T\\
\frac{\pi\epsilon^4 \alpha^2_{\rm EM} \cos^4\theta_W}{4320m^2_{Z^\prime,0}}\frac{s^2(7s-50m^2_{Z^\prime,0}+64m^2_W)}{m^4_W m^2_{Z^\prime,0}}\,\,\,{\rm for}\,\,m_W < T\,\,{\rm and}\,\,s\simeq 9T^2.
\end{cases}
\end{eqnarray}
\item \underline{${\rm SM}{\rm SM} \rightarrow \chi \bar{\chi}$:} These processes are 
mediated by $s$-channel $Z^\prime$-mediated diagrams and can in principle, 
thermalise the DM $\chi$ with the SM unless they are much smaller than $H$. 
The relevant cross-sections are:
\begin{eqnarray}
\sigma_{f\bar{f} \rightarrow \chi\bar{\chi}} (s) &=& \frac{\epsilon^2 \alpha_{\rm EM} Q^2_f Q^2_\chi g^2_D}{3s}\sqrt{\frac{s-4m^2_\chi}{s-4m^2_f}}\frac{(s+2m^2_f)(s+2m^2_\chi)}{(s-m^2_{Z^\prime,0})^2+m^2_{Z^\prime,0}\Gamma^2_{Z^\prime,0}},\\
\sigma_{W^+W^- \rightarrow \chi\bar{\chi}} (s) &=& \frac{\epsilon^2 \alpha_{\rm EM}\cos^2\theta_W Q^2_\chi g^2_D}{108\,m^4_{Z^\prime,0}}\sqrt{\frac{s-4m^2_\chi}{s-4m^2_W}}\left(1+\frac{2m^2_\chi}{s}\right) \frac{1}{(s-m^2_{Z^\prime,0})^2+m^2_{Z^\prime,0}\Gamma^2_{Z^\prime}} \nonumber\\
&& \times \bigg( 4m^4_W(5s+12m^2_{Z^\prime,0})-16m^6_W+s(s^2+16m^2_{Z^\prime,0}s-34m^4_{Z^\prime,0}) \nonumber\\
&& -2m^2_W(4s^2+11m^2_{Z^\prime,0}s+40m^4_{Z^\prime,0})\bigg).
\end{eqnarray}
From these one can calculate the DM freeze-in abundance from SM contributions 
as follows~\cite{Hall:2009bx}:
\begin{eqnarray}
Y^{\rm SM}_\chi &=& \frac{2025}{\pi^4}\sqrt{\frac{2\pi^2}{45}}\frac{M_P}{g_{*,S}\sqrt{g_{*,\rho}}} \int_{T}^{T_R} \frac{dT}{T^6} \langle \sigma v \rangle_{{\rm SM}\,{\rm SM}\rightarrow \chi\bar{\chi}} n^2_{\rm SM}(T),
\end{eqnarray}
with 
\begin{eqnarray}
\langle \sigma v \rangle_{{\rm SM}\,{\rm SM}\rightarrow \chi\bar{\chi}} &=& \frac{1}{8m^4_{\rm SM}T}\int_{4m^2_\chi}^\infty ds s^{1/2} (s-4m^2_{\rm SM}) \sigma_{{\rm SM}\,{\rm SM}\rightarrow \chi\bar{\chi}}(s) \frac{K_1(\sqrt{s}/T)}{K^2_2(m_{\rm SM}/T)}.%
\end{eqnarray}
\item \underline{$Z^\prime Z^\prime \rightarrow Z^\prime Z^\prime$:} The scatterings 
between $Z^\prime$s via $\phi$-mediated processes has a typical cross-section:
\begin{eqnarray}
&&\sigma_{Z^\prime Z^\prime \rightarrow Z^\prime Z^\prime}v = \frac{g^4_D\,m^4_{Z^\prime,0}}{4608\pi s^{3/2}}\sqrt{s-4m^2_{Z^\prime,0}} \left[4+\frac{(s-2m^2_{Z^\prime,0})^2}{m^4_{Z^\prime,0}} + \frac{(s-2m^2_{Z^\prime,0})^4}{16 m^8_{Z^\prime,0}} \right]\times \nonumber\\ 
&& \bigg[\frac{2}{m^2_\phi (s+m^2_\phi-4m^2_{Z^\prime,0})}  + \frac{1}{(s-m^2_\phi)^2+m^2_\phi \Gamma^2_\phi}\nonumber\\ 
&& + \frac{4(3m^2_\phi-4m^2_{Z^\prime,0})(s-m^2_\phi)-m^2_\phi \Gamma^2_\phi}{(s-4m^2_{Z^\prime,0})(s+2m^2_\phi-4m^2_{Z^\prime,0})\left[(s-m^2_\phi)^2+m^2_\phi\Gamma^2_\phi\right]}\log \left(\frac{m^2_\phi}{s+m^2_\phi-4m^2_{Z^\prime,0}} \right)\bigg].\nonumber\\
\end{eqnarray}
When the corresponding interaction rate is smaller than the 
$H$-rate during the entire $Z^\prime$-dominated era after perturbative 
reheating and prior to the SM radiation dominated era, the $Z^\prime$ do not reach kinetic equilibrium.
\end{enumerate}

\section{DM equation of state}

\begin{figure}
    \centering\includegraphics[width=0.5\linewidth]{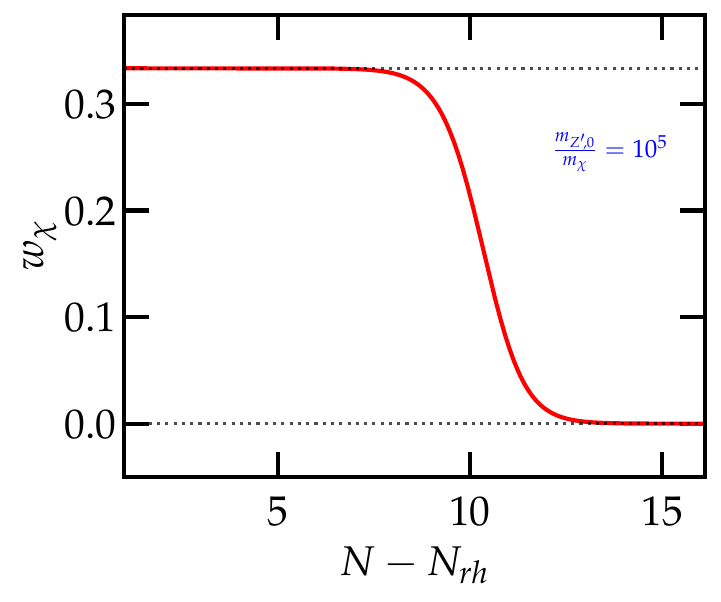}
    \caption{\label{fig:dmeos} The DM equation of state following the end of reheating, using (\ref{eq:chidist}).}
\end{figure}

We consider the non-instantaneous freeze-in production of dark matter (the alternative is inaccurate as $\Gamma \ll H$ in this scenario) by directly solving the Boltzmann equation for the DM phase space distribution. Approximating the non-relativistic $Z'$ reheatons as at rest, neglecting DM-DM scatterings, and following Ref.~\cite{Garcia_2018}, we have
\begin{equation}
    \frac{\partial f_\chi}{\partial t} - Hp\frac{\partial f_\chi}{\partial p} = \frac{2\pi^2}{p^2}n_{Z^\prime}\Gamma_{Z^\prime \rightarrow \chi\bar{\chi}}\delta\left(p - \frac{m_{Z^\prime,0}}{2}\right)
\end{equation}
which has the solution for $t \in [t_{m.d.}, t_{rh}]$
\begin{equation}
    f_\chi (t,p) = \frac{16\pi^2 \Gamma_{Z^\prime \rightarrow \chi\bar{\chi}}}{m_{Z^\prime,0}^3} \int_{t_{\text{m.d.}}}^{t_\text{reh}} \mathrm{d}t'\frac{n_{Z^\prime}}{H}\delta(t' - t_0) ,\quad \text{where}\quad a(t_0) = \frac{2p}{m_{Z^\prime,0}} a(t),
\end{equation}
where we assume negligible production from $t < t_{m.d.}$ for simplicity. Using
\begin{equation}
    n_{Z^\prime}(t_0) = n_{Z^\prime}(t)\left(\frac{m_{Z^\prime,0}}{2p} \right)^3,\quad H(t_0) = H(t>t_{\text{m.d.}}) \left(\frac{m_{Z^\prime,0}}{2p} \right)^{3/2}, \quad \rho_{Z^\prime} \sim m_{Z^\prime,0}n_{Z^\prime}
\end{equation}
we then have
\begin{equation}\label{eq:chidist}
\begin{split}
    f_\chi (t_{\text{m.d.}} < t \leq t_{\text{reh}},p) &\simeq 16\pi^2 \frac{\rho_{Z^\prime}}{m_{Z^\prime,0}^4}\frac{\Gamma_{Z^\prime \rightarrow \chi \bar{\chi}}}{H} \left(\frac{m_{Z^\prime,0}}{2p} \right)^{3/2} \theta(m_{Z^\prime,0} - 2p) 
    \\
    f_\chi (t > t_{\text{reh}},p) &\simeq 16\pi^2 \left[\frac{\rho_{Z^\prime}}{m_{Z^\prime,0}^4}\frac{\Gamma_{Z^\prime \rightarrow \chi \bar{\chi}}}{H}\right]_{t=t_{\text{rh}}} \left(\frac{m_{Z^\prime,0}}{2p} \right)^{3/2} \left(\frac{a_{\text{rh}}}{a} \right)^{3/2} \theta\left(\frac{a_{\text{rh}}}{a}m_{Z^\prime,0} - 2p\right).
\end{split}
\end{equation}
After reheating ends, we have neglected thermal DM production so that the distribution function simply redshifts to the present.
Then, we can obtain $\rho_\chi$ and $P_\chi$ after integration which together give the equation of state for the DM ($w_\chi = P_\chi/\rho_\chi$), plotted in Figure \ref{fig:dmeos} for $m_\chi/m_{Z^\prime,0} = 10^{-5}$, demonstrating that $w\rightarrow 0$ after sufficient expansion. 
\label{sec:appDM}


\providecommand{\href}[2]{#2}
\bibliographystyle{JHEP}
\bibliography{VDMreheating}

\end{document}